\documentclass[12pt]{article}
\usepackage{amsmath,amssymb}
\usepackage{graphicx}
\newcommand{\eqdef}{\stackrel{\text{def}}{=}}
\newcommand{\eqdefrm}{\stackrel{\text{\rm def}}{=}}
\newcommand{\n}{\nonumber\\}
\newcommand{\bm}{\boldsymbol}
\newcommand{\ignore}[1]{}
\newcommand{\Romannumeral}[1]{\uppercase\expandafter{\romannumeral#1}}

\newcommand{\III}{\text{\Romannumeral{3}}}
\newtheorem{defi}{\bf Definition}
\newtheorem{thm}{\bf Theorem}
\newtheorem{cor}{\bf Corollary}
\newcounter{mybangou}

\newtheorem{mylemma}{\bf Lemma}[mybangou]
\newtheorem{myprop}{\bf Proposition}[mybangou]

\newcommand{\myclearandadd}{
  \setcounter{mydefi}{0}
  \setcounter{mylemma}{0}
  \setcounter{myprop}{0}
  \setcounter{mythm}{0}
  \setcounter{mycor}{0}
  \setcounter{myconj}{0}
  \addtocounter{mybangou}{1}
}
\newcommand{\heihoukon}{\raisebox{1mm}{$\sqrt{~~~}\,$}}

\allowdisplaybreaks[4]

\begin{document}

\baselineskip=20pt

\newcommand{\preprint}{
\vspace*{-20mm}
   \begin{flushright}\normalsize \sf
    DPSU-20-1\\
  \end{flushright}}
\newcommand{\Title}[1]{{\baselineskip=26pt
  \begin{center} \Large \bf #1 \\ \ \\ \end{center}}}
\newcommand{\Author}{\begin{center}
  \large \bf Satoru Odake \end{center}}
\newcommand{\Address}{\begin{center}
     Faculty of Science, Shinshu University,
     Matsumoto 390-8621, Japan
   \end{center}}
\newcommand{\Accepted}[1]{\begin{center}
  {\large \sf #1}\\ \vspace{1mm}{\small \sf Accepted for Publication}
  \end{center}}

\preprint
\thispagestyle{empty}

\Title{Wronskian/Casoratian Identities\\ and\\
their Application to Quantum Mechanical Systems}

\Author

\Address
\vspace{1cm}

\begin{abstract}
Corresponding to a certain Wronskian identity, we present two types of new
Casoratian identities.
We apply these identities to the Darboux transformations of quantum mechanical
systems. The Wronskian identity is applied to the ordinary quantum mechanics,
and the two Casoratian identities are applied to the discrete quantum mechanics
with pure imaginary and real shifts, respectively.
\end{abstract}

\section{Introduction}
\label{sec:intro}

The Wronski determinant, Wronskian, is a useful tool for analysis.
For example, the linear independence of $n$ functions can be checked by
calculating their Wronskian.
The Wronskian is a determinant of derivatives of functions. Its difference
version is the Casorati determinant, Casoratian. Corresponding to the type of
difference operations, there are several types of Casoratians.
The Wronskian and Casoratians appear in the study of quantum mechanical
systems, especially for the deformations by multi-step Darboux transformations.

We have considered three types of quantum mechanical systems: oQM, idQM and
rdQM \cite{os24}.
Based on them we have studied the new type of orthogonal polynomials,
exceptional or multi-indexed polynomials
\cite{gomez}--\cite{idQMcH}.
The Schr\"odinger equation is a second order differential equation for
ordinary quantum mechanics (oQM), and a second order difference equation for
discrete quantum mechanics (dQM).
Discrete quantum mechanics with pure imaginary shifts (idQM) is dQM for
the continuous coordinate, and discrete quantum mechanics with real shifts
(rdQM) is dQM for the discrete coordinate.
In our study of deformations of these systems by multi-step Darboux
transformations \cite{gos,os22,os25,os26,os27,os35,idQMcH},
the following Wronskian and Casoratian identities ($n\geq 0$)
have played a very important role (See \S\,\ref{sec:WCid} for definitions of
Casoratians $\text{W}_{\gamma}$ and $\text{W}_{\text{C}}$):
\begin{align}
  \text{oQM}:&\ \ \text{W}\bigl[\text{W}[f_1,f_2,\ldots,f_n,g],
  \text{W}[f_1,f_2,\ldots,f_n,h]\,\bigr](x)\n
  &\ \ =\text{W}[f_1,f_2,\ldots,f_n](x)\,
  \text{W}[f_1,f_2,\ldots,f_n,g,h](x),
  \label{Wformula}\\
  \text{idQM}:&\ \ \text{W}_{\gamma}\bigl[
  \text{W}_{\gamma}[f_1,f_2,\ldots,f_n,g],
  \text{W}_{\gamma}[f_1,f_2,\ldots,f_n,h]\,\bigr](x)\n
  &\ \ =\text{W}_{\gamma}[f_1,f_2,\ldots,f_n](x)\,
  \text{W}_{\gamma}[f_1,f_2,\ldots,f_n,g,h](x),
  \label{idQM:Cformula}\\
  \text{rdQM}:&\ \ \text{W}_{\text{C}}\bigl[
  \text{W}_{\text{C}}[f_1,f_2,\ldots,f_n,g],
  \text{W}_{\text{C}}[f_1,f_2,\ldots,f_n,h]\,\bigr](x)\n
  &\ \ =\text{W}_{\text{C}}[f_1,f_2,\ldots,f_n](x+1)\,
  \text{W}_{\text{C}}[f_1,f_2,\ldots,f_n,g,h](x).
  \label{rdQM:Cformula}
\end{align}

There is a nice generalization of the Wronskian identity \eqref{Wformula}
\cite{swia}. It is Theorem\,\ref{thm:QM} \eqref{QM:thm}, and the above
identity corresponds to $m=2$ case.
It is expected that Casoratian identities \eqref{idQM:Cformula} and
\eqref{rdQM:Cformula} have also similar generalizations.
The first purpose of this paper is to find Casoratian identities corresponding
to Theorem\,\ref{thm:QM}.
They are presented as Theorem\,\ref{thm:idQM} and \ref{thm:rdQM}. 

The second purpose of this paper is the application of
Theorem\,\ref{thm:QM}--\ref{thm:rdQM}.
We apply them to the deformation of quantum mechanical systems by
multi-step Darboux transformations.
We consider quantum mechanical systems, whose Schr\"odinger equation is
\eqref{H:Scheq}.
For any solution $\tilde{\phi}(x)$ of the Schr\"odinger equation
$\mathcal{H}\tilde{\phi}(x)=\tilde{\mathcal{E}}\tilde{\phi}(x)$, which may
not belong to the Hilbert space (namely, may not be square integrable),
the Hamiltonian can be written as
$\mathcal{H}=\hat{\mathcal{A}}^{\dagger}\hat{\mathcal{A}}+\tilde{\mathcal{E}}$,
where $\hat{\mathcal{A}}$ is some operator depending on $\tilde{\phi}$ and
satisfies $\hat{\mathcal{A}}\tilde{\phi}(x)=0$ (some modification is needed
for rdQM).
The Darboux transformation with the seed solution $\tilde{\phi}$ maps the
Hamiltonian $\mathcal{H}$ to $\mathcal{H}^{\text{new}}=\hat{\mathcal{A}}
\hat{\mathcal{A}}^{\dagger}+\tilde{\mathcal{E}}$, and the transformed
eigenfunctions $\phi^{\text{new}}_n(x)=\hat{\mathcal{A}}\phi_n(x)$ satisfy
$\mathcal{H}^{\text{new}}\phi^{\text{new}}_n(x)
=\mathcal{E}_n\phi^{\text{new}}_n(x)$.
The operators $\hat{\mathcal{A}}$ and $\hat{\mathcal{A}}^{\dagger}$ may have
zero modes (in the Hilbert space). For example, when the eigenfunction
$\phi_n$ is taken as a seed solution, $\hat{\mathcal{A}}$ has a zero mode,
$\hat{\mathcal{A}}\phi_n(x)=0$.
Therefore, a Darboux transformation deforms a system almost isospectrally.
The property of the deformed system depends on the employed seed solution:
\begin{equation*}
\begin{tabular}{ccc}
  seed solution&&deformed system\\
  virtual state wavefuntion&$\Rightarrow$&isospectral\\
  eigenstate wavefunction&$\Rightarrow$&state deleted\\
  pseudo virtual state wavefunction&$\Rightarrow$&\ state added\ .
\end{tabular}
\end{equation*}
(Both virtual and pseudo virtual states do not belong to the Hilbert space.
For a virtual state, both $\hat{\mathcal{A}}$ and
$\hat{\mathcal{A}}^{\dagger}$ have no zero mode.
For a pseudo virtual state, $\hat{\mathcal{A}}$ has no zero mode but
$\hat{\mathcal{A}}^{\dagger}$ has a zero mode with new energy eigenvalue.
See \cite{os29} for more explicit conditions for (pseudo) virtual state in oQM.)
When the eigenfunctions of the original system are described by the orthogonal
polynomial $P_n$, those of the deformed system are described by the
multi-indexed orthogonal polynomials $P_{\mathcal{D},n}$, where $\mathcal{D}$
is the set of the labels of the seed solutions.
The characteristic feature of the multi-indexed orthogonal polynomials is
the missing of degrees.
When the set of missing degrees $\mathcal{I}=\mathbb{Z}_{\geq 0}\backslash
\{\deg P_{\mathcal{D},n}|n\in\mathbb{Z}_{\geq 0}\}$ is
$\mathcal{I}=\{0,1,\ldots,\ell-1\}$ ($\ell$ : a positive integer),
we call $P_{\mathcal{D},n}$ a case-(1) multi-indexed polynomial,
and otherwise we call it a case-(2) polynomial.
The situation of case-(1) is called stable in \cite{gkm11}.
When only the virtual state wavefunctions are used as seed solutions,
the case-(1) multi-indexed polynomials are obtained, and in the other
combinations, the case-(2) multi-indexed polynomials are obtained.
We consider the multi-step Darboux transformations using both virtual state
wavefunctions labeled by $\mathcal{D}_{\text{v}}$ and eigenstate wavefunctions
labeled by $\mathcal{D}_{\text{e}}$ as seed solutions.
In this case, no state with new energy eigenvalue is added.
We interpret this in two ways:
\begin{equation}
  \begin{array}{cl}
  (\text{\romannumeral1})&\!\!:\ \mathcal{H}
  \xrightarrow{\ \ \ \ \ \ \ \,\text{virtual states and eigenstates of
  $\mathcal{H}$}\ \ \ \ \ \ \ \ }
  \mathcal{H}_{\mathcal{D}_{\text{v}}\cup\mathcal{D}_{\text{e}}}\\[4pt]
  (\text{\romannumeral2})&\!\!:\ \mathcal{H}
  \xrightarrow{\ \text{virtual states of $\mathcal{H}$}\ }
  \mathcal{H}_{\mathcal{D}_{\text{v}}}
  \xrightarrow{\ \text{eigenstates of
  $\mathcal{H}_{\mathcal{D}_{\text{v}}}$}\ }
  \mathcal{H}_{\mathcal{D}_{\text{v}}\cup\mathcal{D}_{\text{e}}}
  \end{array}.
  \label{twodeform}
\end{equation}
The first interpretation (\romannumeral1) is straightforward.
The second one (\romannumeral2) consists of two steps.
After deforming the original Hamiltonian $\mathcal{H}$ by the Darboux
transformations with only the virtual state wavefunctions as seed solutions,
we deform the deformed Hamiltonian
$\mathcal{H}_{\mathcal{D}_{\text{v}}}$ by the Darboux transformations
with the eigenstate wavefunctions of $\mathcal{H}_{\mathcal{D}_{\text{v}}}$
as seed solutions.
Corresponding to these two interpretations, the eigenfunctions of the deformed
Hamiltonian $\mathcal{H}_{\mathcal{D}_{\text{v}}\cup\mathcal{D}_{\text{e}}}$
are expressed in two ways, and they should agree.
The agreement of these two expressions is shown by using the Wronskian and
Casoratian identities Theorem\,\ref{thm:QM}--\ref{thm:rdQM}.

This paper is organized as follows.
In section \ref{sec:WCid} the Wronskian identities are recapitulated and
two types of the Casoratian identities are presented.
In section \ref{sec:app}, Theorem\,\ref{thm:QM}, \ref{thm:idQM} and
\ref{thm:rdQM} are applied to quantum mechanical systems, oQM, idQM and rdQM,
respectively.
Section \ref{sec:summary} is for a summary and comments.

\section{Wronskian and Casoratian Identities}
\label{sec:WCid}

In this section, after recapitulating the known Wronskian identities,
we derive two types of Casoratian identities.
To the best of our knowledge, Theorem\,\ref{thm:idQM} and \ref{thm:rdQM} are
new results.

\subsection{Wronskian identities}
\label{sec:Wid}
\myclearandadd

In our study of the deformations of oQM systems \cite{gos,os25},
the Wronskian identity \eqref{Wformula} has been used extensively.
This identity \eqref{Wformula} has an interesting generalization \cite{swia},
Theorem\,\ref{thm:QM}, whose $m=2$ case corresponds to \eqref{Wformula}.
We present the definition of the Wronskian, its basic properties
and Theorem\,\ref{thm:QM}, which is proved in \cite{swia}.
We also present its Corollary.

\begin{defi}
\label{def:QM:Wdef}
The Wronski determinant of a set of $n$ functions $\{f_k(x)\}_{k=1}^n$,
$\text{\rm W}$, is defined by
\begin{equation}
  \text{\rm W}[f_1,\ldots,f_n](x)
  \eqdefrm\det\Bigl(\frac{d^{j-1}f_k(x)}{dx^{j-1}}\Bigr)_{1\leq j,k\leq n},
\end{equation}
(for $n=0$, we set $\text{\rm W}[\cdot](x)=1$).
\end{defi}

\begin{mylemma}
\label{lem:QM:1}
For functions $f(x)$ and $g(x)$,
\begin{equation}
  \frac{d}{dx}\frac{f(x)}{g(x)}
  =\frac{\text{\rm W}[g,f](x)}{g(x)^2}.
  \label{QM:lem1}
\end{equation}
\end{mylemma}

\begin{mylemma}
\label{lem:QM:2}
For functions $f_1(x),\ldots,f_n(x)$ $(n\geq 0)$,
\begin{equation}
  \text{\rm W}[1,f_1,\ldots,f_n](x)
  =\text{\rm W}[f'_1,\ldots,f'_n](x),
  \label{QM:lem2}
\end{equation}
where $f'_k(x)\eqdefrm\frac{d}{dx}f_k(x)$.
\end{mylemma}

\begin{myprop}
\label{prop:QM:1}
For functions $f_1(x),\ldots,f_n(x)$ and $g(x)$ $(n\geq 0)$,
\begin{equation}
  \text{\rm W}[gf_1,\ldots,gf_n](x)
  =\bigl(g(x)\bigr)^n\,
  \text{\rm W}[f_1,\ldots,f_n](x).
  \label{QM:prop1}
\end{equation}
\end{myprop}

\begin{myprop}
\label{prop:QM:2}
For functions $f_1(x),\ldots,f_n(x)$ and $g(x)$ $(n\geq 0)$,
\begin{equation}
  \text{\rm W}[g,f_1,\ldots,f_n](x)
  =\bigl(g(x)\bigr)^{1-n}\,
  \text{\rm W}\bigl[
  \text{\rm W}[g,f_1],\ldots,
  \text{\rm W}[g,f_n]\bigr](x).
  \label{QM:prop2}
\end{equation}
\end{myprop}

\begin{thm}
\label{thm:QM}
$\!\!$\cite{swia}
For functions $f_1(x),\ldots,f_n(x)$ and $u_1(x),\ldots,u_m(x)$
$(n\geq 0,\,m\geq 1)$,
\begin{align}
  &\quad
  \bigl(\text{\rm W}[f_1,\ldots,f_n](x)\bigr)^{m-1}\,
  \text{\rm W}[f_1,\ldots,f_n,u_1,\ldots,u_m](x)\n
  &=\text{\rm W}\bigl[
  \text{\rm W}[f_1,\ldots,f_n,u_1],\ldots,
  \text{\rm W}[f_1,\ldots,f_n,u_m]\bigr](x).
  \label{QM:thm}
\end{align}
\end{thm}
This theorem is proved by induction on $n$.
By applying Proposition\,\ref{prop:QM:1} to Theorem\,\ref{thm:QM}
(for later use, $n$ is changed to $l$),
we obtain the following.
\begin{cor}
\label{cor:QM}
For functions $f_1(x),\ldots,f_l(x)$ and $u_1(x),\ldots,u_m(x)$
$(l\geq 0,\,m\geq 1)$,
\begin{equation}
  \frac{\text{\rm W}[f_1,\ldots,f_l,u_1,\ldots,u_m](x)}
  {\text{\rm W}[f_1,\ldots,f_l](x)}
  =\text{\rm W}\biggl[
  \frac{\text{\rm W}[f_1,\ldots,f_l,u_1]}
  {\text{\rm W}[f_1,\ldots,f_l]},\ldots,
  \frac{\text{\rm W}[f_1,\ldots,f_l,u_m]}
  {\text{\rm W}[f_1,\ldots,f_l]}\biggr](x).
  \label{QM:cor}
\end{equation}
\end{cor}

\subsection{Casoratian identities for idQM}
\label{sec:Caiid}
\myclearandadd

Next let us consider the Casoratian appearing in idQM.
In our study of the deformations of idQM systems \cite{gos,os27,idQMcH},
the Casoratian identity \eqref{idQM:Cformula} has been used extensively.
Parallel to the Wronskian in \S\,\ref{sec:Wid},
we present the definition of the Casoratian, its basic properties,
Theorem and Corollary. Here we present their proofs.
We use the convention $\prod_{j=n}^{n-1}a_j=1$.

\begin{defi}
\label{def:idQM:Wdef}
The Casorati determinant of a set of $n$ functions $\{f_k(x)\}_{k=1}^n$,
$\text{\rm W}_{\gamma}$, is defined by
\begin{equation}
  \text{\rm W}_{\gamma}[f_1,\ldots,f_n](x)
  \eqdefrm i^{\frac12n(n-1)}
  \det\Bigl(f_k\bigl(x^{(n)}_j\bigr)\Bigr)_{1\leq j,k\leq n},\quad
  x_j^{(n)}\eqdefrm x+i(\tfrac{n+1}{2}-j)\gamma,
\end{equation}
(for $n=0$, we set $\text{\rm W}_{\gamma}[\cdot](x)=1$).
Here $\gamma$ is a nonzero real constant and $i$ is the imaginary unit.
\end{defi}

\begin{mylemma}
\label{lem:idQM:1}
For functions $f(x)$ and $g(x)$,
\begin{equation}
  \frac{f(x-i\frac{\gamma}{2})}{g(x-i\frac{\gamma}{2})}
  -\frac{f(x+i\frac{\gamma}{2})}{g(x+i\frac{\gamma}{2})}
  =\frac{\text{\rm W}_{\gamma}[g,f](x)}
  {ig(x-i\frac{\gamma}{2})g(x+i\frac{\gamma}{2})}.
  \label{idQM:lem1}
\end{equation}
\end{mylemma}
Proof: Direct calculation shows this lemma.
\hfill\fbox{}

\begin{mylemma}
\label{lem:idQM:2}
For functions $f_1(x),\ldots,f_n(x)$ $(n\geq 0)$,
\begin{equation}
  \text{\rm W}_{\gamma}[1,f_1,\ldots,f_n](x)
  =i^n\text{\rm W}_{\gamma}[Df_1,\ldots,Df_n](x),
  \label{idQM:lem2}
\end{equation}
where $Df_k(x)\eqdefrm f_k(x-i\frac{\gamma}{2})-f_k(x+i\frac{\gamma}{2})$.
\end{mylemma}
Proof: By definition, the LHS is written as a determinant.
In the determinant, subtract the $j$-th row from the $(j+1)$-th row
($j=n,\ldots,2,1$ in turn), and expand the determinant along the $1$-st column.
Since $x^{(n+1)}_{j+1}=x^{(n)}_j-i\frac{\gamma}{2}$ and
$x^{(n+1)}_j=x^{(n)}_j+i\frac{\gamma}{2}$, we obtain the RHS.
\hfill\fbox{}

\noindent
Remark: The LHS of \eqref{idQM:lem1} is expressed as $D\,\frac{f}{g}\,(x)$.

\begin{myprop}
\label{prop:idQM:1}
For functions $f_1(x),\ldots,f_n(x)$ and $g(x)$ $(n\geq 0)$,
\begin{equation}
  \text{\rm W}_{\gamma}[gf_1,\ldots,gf_n](x)
  =\prod_{j=1}^ng\bigl(x^{(n)}_j\bigr)\cdot
  \text{\rm W}_{\gamma}[f_1,\ldots,f_n](x).
  \label{idQM:prop1}
\end{equation}
\end{myprop}
Proof: By definition, the LHS is written as a determinant.
In the determinant, for each $j$-th row, move the factor $g(x^{(n)}_j)$
out of the determinant.
\hfill\fbox{}

\begin{myprop}
\label{prop:idQM:2}
For functions $f_1(x),\ldots,f_n(x)$ and $g(x)$ $(n\geq 0)$,
\begin{equation}
  \text{\rm W}_{\gamma}[g,f_1,\ldots,f_n](x)
  =g\bigl(x^{(n+1)}_1\bigr)
  \prod_{j=1}^n\frac{1}{g(x^{(n+1)}_j)}\cdot
  \text{\rm W}_{\gamma}\bigl[
  \text{\rm W}_{\gamma}[g,f_1],\ldots,
  \text{\rm W}_{\gamma}[g,f_n]\bigr](x).
  \label{idQM:prop2}
\end{equation}
\end{myprop}
Remark: The overall factor in the RHS is written as
$\prod_{j=2}^ng(x^{(n+1)}_j)^{-1}$ for $n\geq 1$.\\
Proof: 
\begin{align*}
  \text{LHS}&\stackrel{(\text{\romannumeral1})}{=}
  \prod_{j=1}^{n+1}g\bigl(x^{(n+1)}_j\bigr)\cdot
  \text{W}_{\gamma}\bigl[1,\tfrac{f_1}{g},\cdots,\tfrac{f_n}{g}\bigr](x)
  \stackrel{(\text{\romannumeral2})}{=}
  \prod_{j=1}^{n+1}g\bigl(x^{(n+1)}_j\bigr)\cdot i^n
  \text{W}_{\gamma}\bigl[D\tfrac{f_1}{g},\cdots,D\tfrac{f_n}{g}\bigr](x)\\
  &\stackrel{(\text{\romannumeral3})}{=}
  \prod_{j=1}^{n+1}g\bigl(x^{(n+1)}_j\bigr)\cdot i^n
  \prod_{j=1}^n\frac{1}{ig(x^{(n)}_j-i\frac{\gamma}{2})
  g(x^{(n)}_j+i\frac{\gamma}{2})}\cdot
  \text{W}_{\gamma}\bigl[
  \text{W}_{\gamma}[g,f_1],\ldots,\text{W}_{\gamma}[g,f_n]\bigr](x)\\
  &\stackrel{(\text{\romannumeral4})}{=}
  \text{RHS},
\end{align*}
where we have used (\romannumeral1): Proposition\,\ref{prop:idQM:1},
(\romannumeral2): Lemma\,\ref{lem:idQM:2},
(\romannumeral3): Lemma\,\ref{lem:idQM:1} (with the remark below
Lemma\,\ref{lem:idQM:2}) and Proposition\,\ref{prop:idQM:1},
(\romannumeral4): $x^{(n)}_j-i\frac{\gamma}{2}=x^{(n+1)}_{j+1}$ and
$x^{(n)}_j+i\frac{\gamma}{2}=x^{(n+1)}_j$.
\hfill\fbox{}

\medskip

The following theorem is a new result.
\begin{thm}
\label{thm:idQM}
For functions $f_1(x),\ldots,f_n(x)$ and $u_1(x),\ldots,u_m(x)$
$(n\geq 0,\,m\geq 1)$,
\begin{align}
  &\quad
  \prod_{j=1}^{m-1}\text{\rm W}_{\gamma}[f_1,\ldots,f_n]
  \bigl(x^{(m-1)}_j\bigr)\cdot
  \text{\rm W}_{\gamma}[f_1,\ldots,f_n,u_1,\ldots,u_m](x)\n
  &=\text{\rm W}_{\gamma}\bigl[
  \text{\rm W}_{\gamma}[f_1,\ldots,f_n,u_1],\ldots,
  \text{\rm W}_{\gamma}[f_1,\ldots,f_n,u_m]\bigr](x).
  \label{idQM:thm}
\end{align}
\end{thm}
Proof: Let us prove this theorem by induction on $n$.
It is trivial for $n=0$.
For $n>0$, since it is trivial for $f_1(x)=0$, we assume $f_1(x)\neq 0$.
For $n=1$, we have
\begin{align*}
  &\quad\prod_{j=1}^{m-1}\text{W}_{\gamma}[f_1]\bigl(x^{(m-1)}_j\bigr)\cdot
  \text{W}_{\gamma}[f_1,u_1,\ldots,u_m](x)\n
  &\stackrel{(\text{\romannumeral1})}{=}
  \prod_{j=1}^{m-1}f_1\bigl(x^{(m-1)}_j\bigr)\cdot
  \prod_{j=2}^m\frac{1}{f_1\bigl(x^{(m+1)}_j\bigr)}\cdot
  \text{W}_{\gamma}\bigl[
  \text{W}_{\gamma}[f_1,u_1],\ldots,
  \text{W}_{\gamma}[f_1,u_m]\bigr](x)\n
  &\stackrel{(\text{\romannumeral2})}{=}
  \text{W}_{\gamma}\bigl[
  \text{W}_{\gamma}[f_1,u_1],\ldots,
  \text{W}_{\gamma}[f_1,u_m]\bigr](x),
\end{align*}
where we have used (\romannumeral1): Proposition\,\ref{prop:idQM:2},
(\romannumeral2): $x^{(m-1)}_j=x^{(m+1)}_{j+1}$.
Hence $n=1$ case holds.\\
Assume that \eqref{idQM:thm} holds till $n$ ($n\geq1$).
Then we have
\begin{align*}
  &\quad\text{W}_{\gamma}\bigl[
  \text{W}_{\gamma}[f_1,\ldots,f_{n+1},u_1],\ldots,
  \text{W}_{\gamma}[f_1,\ldots,f_{n+1},u_m]\bigr](x)\\
  &\stackrel{(\text{\romannumeral1})}{=}\text{W}_{\gamma}\Bigl[
  g\text{W}_{\gamma}\bigl[
  \text{W}_{\gamma}[f_1,f_2],\ldots,
  \text{W}_{\gamma}[f_1,f_{n+1}],
  \text{W}_{\gamma}[f_1,u_1]\bigr],\ldots,\\
  &\qquad\quad\ g\text{W}_{\gamma}\bigl[
  \text{W}_{\gamma}[f_1,f_2],\ldots,
  \text{W}_{\gamma}[f_1,f_{n+1}],
  \text{W}_{\gamma}[f_1,u_m]\bigr]\Bigr](x)\qquad
  \Bigl(g(x)\eqdef\prod_{j=2}^{n+1}\frac{1}{f_1\bigl(x^{(n+2)}_j\bigr)}\Bigr)\\
  &\stackrel{(\text{\romannumeral2})}{=}
  \prod_{l=1}^mg\bigl(x^{(m)}_l\bigr)\cdot
  \text{W}_{\gamma}\Bigl[
  \text{W}_{\gamma}\bigl[
  \text{W}_{\gamma}[f_1,f_2],\ldots,
  \text{W}_{\gamma}[f_1,f_{n+1}],
  \text{W}_{\gamma}[f_1,u_1]\bigr],\ldots,\\
  &\hspace{37mm}\text{W}_{\gamma}\bigl[
  \text{W}_{\gamma}[f_1,f_2],\ldots,
  \text{W}_{\gamma}[f_1,f_{n+1}],
  \text{W}_{\gamma}[f_1,u_m]\bigr]\Bigr](x)\\
  &\stackrel{(\text{\romannumeral3})}{=}
  \prod_{l=1}^mg\bigl(x^{(m)}_l\bigr)\cdot
  \prod_{j=1}^{m-1}\text{W}_{\gamma}\bigl[
  \text{W}_{\gamma}[f_1,f_2],\ldots,
  \text{W}_{\gamma}[f_1,f_{n+1}]\bigr]\bigl(x^{(m-1)}_j\bigr)\\
  &\quad\times
  \text{W}_{\gamma}\bigl[
  \text{W}_{\gamma}[f_1,f_2],\ldots,
  \text{W}_{\gamma}[f_1,f_{n+1}],
  \text{W}_{\gamma}[f_1,u_1],\ldots,
  \text{W}_{\gamma}[f_1,u_m]\bigr](x)\\
  &\stackrel{(\text{\romannumeral4})}{=}
  \prod_{l=1}^mg\bigl(x^{(m)}_l\bigr)\cdot
  \prod_{j=1}^{m-1}\Bigl(
  \prod_{l=2}^nf_1\bigl(x^{(m-1)}_j+i(\tfrac{n+2}{2}-l)\gamma\bigr)\cdot
  \text{W}_{\gamma}[f_1,f_2,\ldots,f_{n+1}]\bigr]\bigl(x^{(m-1)}_j\bigr)\Bigr)\\
  &\quad\times
  \prod_{l=2}^{n+m}f_1\bigl(x^{(n+m+1)}_l\bigr)\cdot
  \text{W}_{\gamma}[f_1,f_2,\ldots,f_{n+1},u_1,\ldots,u_m](x)\\
  &\stackrel{(\text{\romannumeral5})}{=}
  \prod_{j=1}^{m-1}\text{W}_{\gamma}[f_1,f_2,\ldots,f_{n+1}]
  \bigl(x^{(m-1)}_j\bigr)
  \cdot\text{W}_{\gamma}[f_1,f_2,\ldots,f_{n+1},u_1,\ldots,u_m](x),
\end{align*}
where we have used (\romannumeral1): Proposition\,\ref{prop:idQM:2},
(\romannumeral2): Proposition\,\ref{prop:idQM:1},
(\romannumeral3): induction assumption,
(\romannumeral4): Proposition\,\ref{prop:idQM:2},
(\romannumeral5): calculation of $f_1$ factors.
Therefore $n+1$ case also holds.
This concludes the induction proof of \eqref{idQM:thm}.
\hfill\fbox{}

\noindent
The Casoratian identity \eqref{idQM:Cformula} corresponds to $m=2$ case of
Theorem\,\ref{thm:idQM}.

\medskip

We present a corollary of Theorem\,\ref{thm:idQM}
(for later use, $n$ is changed to $l$).
\begin{cor}
\label{cor:idQM}
For functions $f_1(x),\ldots,f_l(x)$ and $u_1(x),\ldots,u_m(x)$
$(l\geq 0,\,m\geq 1)$,
\begin{align}
  &\frac{\text{\rm W}_{\gamma}[f_1,\ldots,f_l,u_1,\ldots,u_m](x)}
  {\sqrt{\text{\rm W}_{\gamma}[f_1,\ldots,f_l](x-i\frac{m}{2}\gamma)
  \text{\rm W}_{\gamma}[f_1,\ldots,f_l](x+i\frac{m}{2}\gamma)}}\n
  &\quad=\text{\rm W}_{\gamma}\biggl[
  \frac{\text{\rm W}_{\gamma}[f_1,\ldots,f_l,u_1]}{w},\ldots,
  \frac{\text{\rm W}_{\gamma}[f_1,\ldots,f_l,u_m]}{w}\biggr](x),
  \label{idQM:cor}
\end{align}
where $\displaystyle w(x)\eqdefrm\sqrt{
\text{\rm W}_{\gamma}[f_1,\ldots,f_l](x-i\tfrac{\gamma}{2})
\text{\rm W}_{\gamma}[f_1,\ldots,f_l](x+i\tfrac{\gamma}{2})}$.
\end{cor}
Proof:
\begin{align*}
  \text{RHS}&\stackrel{(\text{\romannumeral1})}{=}
  \prod_{j=1}^m\frac{1}{w(x^{(l)}_j)}\cdot
  \text{W}_{\gamma}\bigl[
  \text{W}_{\gamma}[f_1,\ldots,f_l,u_1],\ldots,
  \text{W}_{\gamma}[f_1,\ldots,f_l,u_m]\bigr](x)\\
  &\stackrel{(\text{\romannumeral2})}{=}
  \prod_{j=1}^m\frac{1}{w(x^{(l)}_j)}\cdot
  \prod_{j=1}^{m-1}\text{\rm W}_{\gamma}[f_1,\ldots,f_l]
  \bigl(x^{(m-1)}_j\bigr)\cdot
  \text{\rm W}_{\gamma}[f_1,\ldots,f_l,u_1,\ldots,u_m](x)\\
  &\stackrel{(\text{\romannumeral3})}{=}
  \text{LHS},
\end{align*}
where we have used (\romannumeral1): Proposition\,\ref{prop:idQM:1},
(\romannumeral2): Theorem\,\ref{thm:idQM},
(\romannumeral3): direct calculation.
\hfill\fbox{}

\noindent
Remark: We regard the square root function $\heihoukon$ in
Corollary\,\ref{cor:idQM} as a complex function.

\medskip

The Casoratian $\text{W}_{\gamma}$ reduces to the Wronskian $\text{W}$
in the $\gamma\to 0$ limit.
\begin{myprop}
\label{prop:idQM:3}
\begin{equation}
  \lim_{\gamma\to 0}\gamma^{-\frac12n(n-1)}
  \text{\rm W}_{\gamma}[f_1,\ldots,f_n](x)
  =\text{\rm W}[f_1,\ldots,f_n](x).
  \label{idQM:prop3}
\end{equation}
\end{myprop}
Proof: For a determinant of an $n\times n$ matrix, let us define the operation
$O_m$ ($1\leq m\leq n-1$) as follows: Subtract the $j$-th row from the
$(j+1)$-th row ($j=n-1,n-2,\ldots,m$ in turn).
By definition, $\text{W}_{\gamma}[f_1,\ldots,f_n](x)$ is written as a
determinant. Apply the operations $O_m$ ($m=1,2,\ldots,n-1$ in turn) to the
determinant.
Then the $(j,k)$-element of the determinant becomes
\begin{align*}
  &\quad\sum_{r=0}^{j-1}(-1)^r\genfrac{(}{)}{0pt}{}{j-1}{r}
  f_k\big(x^{(n)}_{j-r}\bigr)
  =\sum_{r=0}^{j-1}(-1)^r\genfrac{(}{)}{0pt}{}{j-1}{r}
  f_k\big(x+i\tfrac{n-j}{2}\gamma+i(r-\tfrac{j-1}{2})\gamma\bigr)\\
  &\stackrel{(\text{\romannumeral1})}{=}
  \sum_{r=0}^{j-1}(-1)^r\genfrac{(}{)}{0pt}{}{j-1}{r}
  \Bigl(\sum_{s=0}^{j-1}\frac{1}{s!}
  \frac{d^s}{dx^s}f_k\bigl(x+i\tfrac{n-j}{2}\gamma\bigr)
  \bigl(i(r-\tfrac{j-1}{2})\gamma\bigr)^s+O(\gamma^j)\Bigr)\\
  &=\sum_{s=0}^{j-1}\frac{(i\gamma)^s}{s!}
  \frac{d^s}{dx^s}f_k\bigl(x+i\tfrac{n-j}{2}\gamma\bigr)
  \sum_{r=0}^{j-1}(-1)^r\genfrac{(}{)}{0pt}{}{j-1}{r}
  \bigl(r-\tfrac{j-1}{2}\bigr)^s+O(\gamma^j)\\
  &\stackrel{(\text{\romannumeral2})}{=}
  (-i\gamma)^{j-1}
  \frac{d^{j-1}}{dx^{j-1}}f_k\bigl(x+i\tfrac{n-j}{2}\gamma\bigr)
  +O(\gamma^j)\\
  &=(-i\gamma)^{j-1}\frac{d^{j-1}f_k(x)}{dx^{j-1}}
  \times\bigl(1+O(\gamma)\bigr),
\end{align*}
where we have used (\romannumeral1): Taylor expansion,
(\romannumeral2): the following sum formula
\begin{equation*}
  \sum_{r=0}^{j-1}(-1)^r\genfrac{(}{)}{0pt}{}{j-1}{r}
  \bigl(r-\tfrac{j-1}{2}\bigr)^s=(-1)^{j-1}(j-1)!\,\delta_{s,j-1}
  \ \ (0\leq s\leq j-1).
\end{equation*}
Thus we have
\begin{align*}
  \text{W}_{\gamma}[f_1,\ldots,f_n](x)&=i^{\frac12n(n-1)}
  \det\Bigl((-i\gamma)^{j-1}\frac{d^{j-1}f_k(x)}{dx^{j-1}}
  \times\bigl(1+O(\gamma)\bigr)\Bigr)_{1\leq j,k\leq n}\\
  &=i^{\frac12n(n-1)}\prod_{j=1}^n(-i\gamma)^{j-1}\cdot
  \det\Bigl(\frac{d^{j-1}f_k(x)}{dx^{j-1}}
  \times\bigl(1+O(\gamma)\bigr)\Bigr)_{1\leq j,k\leq n}\\
  &=\gamma^{\frac12n(n-1)}\text{W}[f_1,\ldots,f_n](x)
  \times\bigl(1+O(\gamma)\bigr).
\end{align*}
By multiplying $\gamma^{-\frac12n(n-1)}$ and taking the $\gamma\to 0$ limit,
we obtain \eqref{idQM:prop3}.
\hfill\fbox{}

\noindent
By multiplying appropriate powers of $\gamma$ and taking the $\gamma\to 0$
limit, the properties of the Casoratian $\text{W}_{\gamma}$ presented in
this subsection reduce to those of the Wronskian $\text{W}$ in
\S\,\ref{sec:Wid}.

\subsection{Casoratian identities for rdQM}
\label{sec:Carid}
\myclearandadd

Next let us consider the Casoratian appearing in rdQM.
In our study of the deformations of rdQM systems \cite{os22,os26,os35},
the Casoratian identity \eqref{rdQM:Cformula} has been used extensively.
Parallel to the Wronskian in \S\,\ref{sec:Wid},
we present the definition of the Casoratian, its basic properties,
Theorem and Corollary.
Since their proofs are similar to those of \S\,\ref{sec:Caiid}, we omit them.

\begin{defi}
\label{def:rdQM:Wdef}
The Casorati determinant of a set of $n$ functions $\{f_k(x)\}_{k=1}^n$,
$\text{\rm W}_{\text{\rm C}}$, is defined by
\begin{equation}
  \text{\rm W}_{\text{\rm C}}[f_1,\ldots,f_n](x)
  \eqdefrm\det\Bigl(f_k(x+j-1)\Bigr)_{1\leq j,k\leq n},
\end{equation}
(for $n=0$, we set $\text{\rm W}_{\text{\rm C}}[\cdot](x)=1$).
\end{defi}

\begin{mylemma}
\label{lem:rdQM:1}
For functions $f(x)$ and $g(x)$,
\begin{equation}
  \frac{f(x+1)}{g(x+1)}-\frac{f(x)}{g(x)}
  =\frac{\text{\rm W}_{\text{\rm C}}[g,f](x)}{g(x)g(x+1)}.
  \label{rdQM:lem1}
\end{equation}
\end{mylemma}

\begin{mylemma}
\label{lem:rdQM:2}
For functions $f_1(x),\ldots,f_n(x)$ $(n\geq 0)$,
\begin{equation}
  \text{\rm W}_{\text{\rm C}}[1,f_1,\ldots,f_n](x)
  =\text{\rm W}_{\text{\rm C}}[Df_1,\ldots,Df_n](x),
  \label{rdQM:lem2}
\end{equation}
where $Df_k(x)\eqdefrm f_k(x+1)-f_k(x)$.
\end{mylemma}
Remark: The LHS of \eqref{rdQM:lem1} is expressed as $D\,\frac{f}{g}\,(x)$.

\begin{myprop}
\label{prop:rdQM:1}
For functions $f_1(x),\ldots,f_n(x)$ and $g(x)$ $(n\geq 0)$,
\begin{equation}
  \text{\rm W}_{\text{\rm C}}[gf_1,\ldots,gf_n](x)
  =\prod_{j=1}^ng(x+j-1)\cdot
  \text{\rm W}_{\text{\rm C}}[f_1,\ldots,f_n](x).
  \label{rdQM:prop1}
\end{equation}
\end{myprop}

\begin{myprop}
\label{prop:rdQM:2}
For functions $f_1(x),\ldots,f_n(x)$ and $g(x)$ $(n\geq 0)$,
\begin{equation}
  \text{\rm W}_{\text{\rm C}}[g,f_1,\ldots,f_n](x)
  =g(x)\prod_{j=1}^n\frac{1}{g(x+j-1)}\cdot
  \text{\rm W}_{\text{\rm C}}\bigl[
  \text{\rm W}_{\text{\rm C}}[g,f_1],\ldots,
  \text{\rm W}_{\text{\rm C}}[g,f_n]\bigr](x).
  \label{rdQM:prop2}
\end{equation}
\end{myprop}
Remark: The overall factor in the RHS is written as
$\prod_{j=2}^ng(x+j-1)^{-1}$ for $n\geq 1$.

\medskip

The following theorem is a new result.
\begin{thm}
\label{thm:rdQM}
For functions $f_1(x),\ldots,f_n(x)$ and $u_1(x),\ldots,u_m(x)$
$(n\geq 0,\,m\geq 1)$,
\begin{align}
  &\quad
  \prod_{j=1}^{m-1}\text{\rm W}_{\text{\rm C}}[f_1,\ldots,f_n](x+j)\cdot
  \text{\rm W}_{\text{\rm C}}[f_1,\ldots,f_n,u_1,\ldots,u_m](x)\n
  &=\text{\rm W}_{\text{\rm C}}\bigl[
  \text{\rm W}_{\text{\rm C}}[f_1,\ldots,f_n,u_1],\ldots,
  \text{\rm W}_{\text{\rm C}}[f_1,\ldots,f_n,u_m]\bigr](x).
  \label{rdQM:thm}
\end{align}
\end{thm}

\noindent
The Casoratian identity \eqref{rdQM:Cformula} corresponds to $m=2$ case of
Theorem\,\ref{thm:rdQM}.

\medskip

We present a corollary of Theorem\,\ref{thm:rdQM}
(for later use, $n$ is changed to $l$).
\begin{cor}
\label{cor:rdQM}
For functions $f_1(x),\ldots,f_l(x)$ and $u_1(x),\ldots,u_m(x)$
$(l\geq 0,\,m\geq 1)$,
\begin{align}
  &\frac{\text{\rm W}_{\text{\rm C}}[f_1,\ldots,f_l,u_1,\ldots,u_m](x)}
  {\sqrt{\text{\rm W}_{\text{\rm C}}[f_1,\ldots,f_l](x)
  \text{\rm W}_{\text{\rm C}}[f_1,\ldots,f_l](x+m)}}\n
  &\quad=\text{\rm W}_{\text{\rm C}}\biggl[
  \frac{\text{\rm W}_{\text{\rm C}}[f_1,\ldots,f_l,u_1]}{w},\ldots,
  \frac{\text{\rm W}_{\text{\rm C}}[f_1,\ldots,f_l,u_m]}{w}\biggr](x),
  \label{rdQM:cor}
\end{align}
where $w(x)\eqdefrm\sqrt{
\text{\rm W}_{\text{\rm C}}[f_1,\ldots,f_l](x)
\text{\rm W}_{\text{\rm C}}[f_1,\ldots,f_l](x+1)}$.
\end{cor}
Remark: We regard the square root function $\heihoukon$ in
Corollary\,\ref{cor:rdQM} as a real function.
We have assumed $\text{W}_{\text{\rm C}}[f_1,\ldots,f_l](x)>0$.

\section{Application to Quantum Mechanical Systems}
\label{sec:app}

In this section we consider the application of
Theorem\,\ref{thm:QM}--\ref{thm:rdQM} to the deformation of quantum mechanical
systems by multi-step Darboux transformations.
As quantum mechanical systems, we consider oQM, idQM and rdQM, to which
Theorem\,\ref{thm:QM}, \ref{thm:idQM} and \ref{thm:rdQM} are applied
respectively.
For simplicity of presentation, we assume that rdQM systems are semi-infinite
systems.

We assume that the original system with the Hamiltonian $\mathcal{H}$,
which is hermitian and positive semi-definite, has
the eigenfunctions (eigenstate wavefunctions) $\phi_n(x)$,
\begin{equation}
  \mathcal{H}\phi_n(x)=\mathcal{E}_n\phi_n(x),\quad
  0=\mathcal{E}_0<\mathcal{E}_1<\cdots\ \ (n\in\mathbb{Z}_{\geq 0}),
  \label{H:Scheq}
\end{equation}
and the virtual state wavefunctions $\tilde{\phi}_{\text{v}}(x)$
\cite{os25,os26,os27,os35,idQMcH},
\begin{equation}
  \mathcal{H}\tilde{\phi}_{\text{v}}(x)
  =\tilde{\mathcal{E}}_{\text{v}}\tilde{\phi}_{\text{v}}(x),\quad
  \tilde{\mathcal{E}}_{\text{v}}<0.
  \label{vs:Scheq}
\end{equation}
The virtual state wavefunction has a definite sign for the physical value of
$x$.
As seed solutions of the multi-step Darboux transformations, we take both
the virtual state wavefunctions $\tilde{\phi}_{\text{v}}(x)$
($\text{v}\in\mathcal{D}_{\text{v}}$) and the eigenfunctions $\phi_n(x)$
($n\in\mathcal{D}_{\text{e}}$).
Here $\mathcal{D}_{\text{v}}$ and $\mathcal{D}_{\text{e}}$ are sets of labels
of the virtual states and the eigenstates respectively, and we set them as
\begin{equation}
  \mathcal{D}_{\text{v}}\eqdef\{\text{v}_1,\ldots,\text{v}_{M_{\text{v}}}\}
  \ \ (\text{v}_j\in\mathbb{Z}_{\geq 0}),\quad
  \mathcal{D}_{\text{e}}\eqdef\{e_1,\ldots,e_{M_{\text{e}}}\}
  \ \ (e_j\in\mathbb{Z}_{\geq 0}),
\end{equation}
where $\text{v}_j$'s are mutually distinct and $e_j$'s are mutually distinct.
If there are two types of the virtual states, the label includes the type.
By combining these, we set
\begin{equation}
  \mathcal{D}\eqdef\mathcal{D}_{\text{v}}\cup\mathcal{D}_{\text{e}}
  \eqdef\{d_1,\ldots,d_M\},\quad M\eqdef M_{\text{v}}+M_{\text{e}}.
\end{equation}
(Exactly speaking, the index sets $\mathcal{D}_{\text{v}}$,
$\mathcal{D}_{\text{e}}$ and $\mathcal{D}$ are ordered sets, but we do not
care much about the order, because the deformed Hamiltonians
$\mathcal{H}_{\mathcal{D}_{\text{v}}}$, $\mathcal{H}_{\mathcal{D}_{\text{e}}}$
and $\mathcal{H}_{\mathcal{D}}$ do not depend on the order.)
We set seed solutions as $\psi_j(x)$ ($j=1,\ldots,M$), namely,
$\psi_j(x)=\tilde{\phi}_{\text{v}_k}(x)$ for $d_j=\text{v}_k$ and
$\psi_j(x)=\phi_{e_k}(x)$ for $d_j=e_k$.
By the multi-step Darboux transformations with the seed solutions $\psi_j(x)$
($j\in\mathcal{D}$), the Hamiltonian $\mathcal{H}$ is deformed to
$\mathcal{H}_{\mathcal{D}}$.
The Schr\"odinger equation of the deformed system is
\begin{equation}
  \mathcal{H}_{\mathcal{D}}\phi_{\mathcal{D}\,n}(x)
  =\mathcal{E}_n\phi_{\mathcal{D}\,n}(x)
  \ \ (n\in\mathbb{Z}_{\geq 0}\backslash\mathcal{D}_{\text{e}}).
  \label{HD:Scheq}
\end{equation}
If the Krein-Adler condition \cite{krein,adler},
\begin{equation}
  \prod_{j=1}^{M_{\text{e}}}(m-e_j)\ge0\quad
  (\,\forall m\in\mathbb{Z}_{\geq 0}),
  \label{KAcond}
\end{equation}
is satisfied (it is trivial for $\mathcal{D}_{\text{e}}=\emptyset$),
the norm of $\phi_{\mathcal{D}\,n}(x)$ becomes positive definite,
$(\phi_{\mathcal{D}\,n},\phi_{\mathcal{D}\,n})>0$
$(n\in\mathbb{Z}_{\geq 0}\backslash\mathcal{D}_{\text{e}})$
\cite{krein,adler,gos,os22}.
This condition \eqref{KAcond} means
$\mathcal{D}_{\text{e}}=\{0,1,\ldots,n_0\}\cup\bigcup_{l=1}^L\{j_l,j_l+1\}$
($n_0+1,L,j_l\in\mathbb{Z}_{\geq 0}$,
$n_0+1<j_1$, $j_l+2\leq j_{l+1}$, $n_0+1+2L=M_{\text{e}}$,
$\{0,1,\ldots,n_0\}=\emptyset$ for $n_0=-1$ and
$\bigcup_{l=1}^LA_l=\emptyset$ for $L=0$),
or equivalently,
$\mathcal{D}_{\text{e}}=\{0,1,\ldots,n_0\}\cup\bigcup_{l=1}^{L'}
\{k_l,k_l+1,\ldots,k_l+2r_l-1\}$
($n_0+1,L',k_l,r_l-1\in\mathbb{Z}_{\geq 0}$,
$n_0+1<k_1$, $k_l+2r_l<k_{l+1}$, $n_0+1+\sum_{l=1}^{L'}2r_l=M_{\text{e}}$).
For oQM, it is shown that the deformed Hamiltonian $\mathcal{H}_{\mathcal{D}}$
with \eqref{KAcond} is well-defined and hermitian (for an appropriate range
of the parameters) \cite{krein,adler}.
For dQM, it is conjectured that the deformed Hamiltonian
$\mathcal{H}_{\mathcal{D}}$ with \eqref{KAcond} is well-defined and hermitian
(for an appropriate range of the parameters) \cite{gos,os22}.
This is strongly supported by the positive definiteness of the norm.
It is also supported by numerical calculation for each system.
The eigenfunctions $\phi_{\mathcal{D}\,n}(x)$ are expressed in terms of
the Wronskian/Casoratian.

Let us reinterpret this deformation as \eqref{twodeform}.
First, by the multi-step Darboux transformations with the seed solutions
$\tilde{\phi}_{\text{v}}(x)$ ($\text{v}\in\mathcal{D}_{\text{v}}$),
the Hamiltonian $\mathcal{H}$ is deformed to
$\mathcal{H}_{\mathcal{D}_{\text{v}}}$.
The Schr\"odinger equation of this deformed system is
\begin{equation}
  \mathcal{H}_{\mathcal{D}_{\text{v}}}\phi_{\mathcal{D}_{\text{v}}\,n}(x)
  =\mathcal{E}_n\phi_{\mathcal{D}_{\text{v}}\,n}(x)
  \ \ (n\in\mathbb{Z}_{\geq 0}).
  \label{HDv:Scheq}
\end{equation}
The eigenfunctions $\phi_{\mathcal{D}_{\text{v}}\,n}(x)$ are expressed in
terms of the Wronskian/Casoratian, and the deformed Hamiltonian
$\mathcal{H}_{\mathcal{D}_{\text{v}}}$ is well-defined and hermitian
(for an appropriate range of the parameters).
Second, by the multi-step Darboux transformations with the seed solutions
$\phi_{\mathcal{D}_{\text{v}}\,n}(x)$ ($n\in\mathcal{D}_{\text{e}}$),
the Hamiltonian $\mathcal{H}_{\mathcal{D}_\text{v}}$ is deformed to
$\mathcal{H}_{\mathcal{D}_{\text{v}}\mathcal{D}_{\text{e}}}$.
The Schr\"odinger equation of this deformed system is
\begin{equation}
  \mathcal{H}_{\mathcal{D}_{\text{v}}\mathcal{D}_{\text{e}}}
  \phi_{\mathcal{D}_{\text{v}}\mathcal{D}_{\text{e}}\,n}(x)
  =\mathcal{E}_n\phi_{\mathcal{D}_{\text{v}}\mathcal{D}_{\text{e}}\,n}(x)
  \ \ (n\in\mathbb{Z}_{\geq 0}\backslash\mathcal{D}_{\text{e}}).
  \label{HDvDe:Scheq}
\end{equation}
The eigenfunctions $\phi_{\mathcal{D}_{\text{v}}\mathcal{D}_{\text{e}}\,n}(x)$
are expressed in terms of the Wronskian/Casoratian.
If the Krein-Adler condition \eqref{KAcond} is satisfied, the deformed
Hamiltonian $\mathcal{H}_{\mathcal{D}_{\text{v}}\mathcal{D}_{\text{e}}}$ is
well-defined and hermitian (for an appropriate range of the parameters).
Since two deformed Hamiltonian $\mathcal{H}_{\mathcal{D}}$ and
$\mathcal{H}_{\mathcal{D}_{\text{v}}\mathcal{D}_{\text{e}}}$ should be the
same, two eigenfunctions $\phi_{\mathcal{D}\,n}(x)$ and
$\phi_{\mathcal{D}_{\text{v}}\mathcal{D}_{\text{e}}\,n}(x)$ must be the same
(proportional).
We will show this equality $\phi_{\mathcal{D}\,n}(x)=
\phi_{\mathcal{D}_{\text{v}}\mathcal{D}_{\text{e}}\,n}(x)$
by using the Wronskian/Casoratian identities
Theorem\,\ref{thm:QM}--\ref{thm:rdQM}
(Corollary\,\ref{cor:QM}--\ref{cor:rdQM}).

\subsection{Application to oQM}
\label{sec:app_oQM}

First let us consider oQM.
The virtual states are studied for the exactly solvable systems whose
eigenfunctions are described by the Laguerre and Jacobi polynomials
\cite{os25}.
The Hamiltonian $\mathcal{H}$ of oQM has the following form:
\begin{equation}
  \mathcal{H}=p^2+U(x),
  \label{oQM:H}
\end{equation}
where $x$ is the coordinate and $p$ is the momentum, $p=-i\frac{d}{dx}$.
The deformed Hamiltonian $\mathcal{H}_{\mathcal{D}}$ \eqref{HD:Scheq} is
given by
\begin{equation}
  \mathcal{H}_{\mathcal{D}}=p^2+U_{\mathcal{D}}(x),\quad
  U_{\mathcal{D}}(x)=U(x)-2\partial_x^2\log\bigl|
  \text{W}[\psi_1,\ldots,\psi_M](x)\bigr|,
  \label{oQM:HD}
\end{equation}
and its eigenfunctions $\phi_{\mathcal{D}\,n}(x)$ are given by (for example,
see \S\,2 of \cite{gos} and Appendix A of \cite{rrmiop3})
\begin{equation}
  \phi_{\mathcal{D}\,n}(x)=\frac{\text{W}[\psi_1,\ldots,\psi_M,\phi_n](x)}
  {\text{W}[\psi_1,\ldots,\psi_M](x)}
  \ \ (n\in\mathbb{Z}_{\geq 0}\backslash\mathcal{D}_{\text{e}}).
  \label{oQM:phiDn}
\end{equation}
On the other hand, the eigenfunctions of
$\mathcal{H}_{\mathcal{D}_{\text{v}}}$ \eqref{HDv:Scheq} are given by
\cite{os25}
\begin{equation}
  \phi_{\mathcal{D}_{\text{v}}\,n}(x)=
  \frac{\text{W}[\tilde{\phi}_{\text{v}_1},\ldots,
  \tilde{\phi}_{\text{v}_{M_{\text{v}}}},\phi_n](x)}
  {\text{W}[\tilde{\phi}_{\text{v}_1},\ldots,
  \tilde{\phi}_{\text{v}_{M_{\text{v}}}}](x)}
  \ \ (n\in\mathbb{Z}_{\geq 0}).
  \label{oQM:phiDvn}
\end{equation}
So the eigenfunctions of
$\mathcal{H}_{\mathcal{D}_{\text{v}}\mathcal{D}_{\text{e}}}$
\eqref{HDvDe:Scheq} are expressed as
\begin{equation}
  \phi_{\mathcal{D}_{\text{v}}\mathcal{D}_{\text{e}}\,n}(x)
  =\frac{\text{W}[\phi_{\mathcal{D}_{\text{v}}\,e_1},\ldots,
  \phi_{\mathcal{D}_{\text{v}}\,e_{M_{\text{e}}}},
  \phi_{\mathcal{D}_{\text{v}}\,n}](x)}
  {\text{W}[\phi_{\mathcal{D}_{\text{v}}\,e_1},\ldots,
  \phi_{\mathcal{D}_{\text{v}}\,e_{M_{\text{e}}}}](x)}
  \ \ (n\in\mathbb{Z}_{\geq 0}\backslash\mathcal{D}_{\text{e}}).
  \label{oQM:phiDvDen}
\end{equation}
We will show that two expressions \eqref{oQM:phiDn} and \eqref{oQM:phiDvDen}
are actually identical by using the Wronskian identity,
Corollary\,\ref{cor:QM}.

Corollary\,\ref{cor:QM} with the replacements $m\to m+1$ and $u_{m+1}=v$
becomes
\begin{align*}
  &\quad\frac{\text{W}[f_1,\ldots,f_l,u_1,\ldots,u_m,v](x)}
  {\text{W}[f_1,\ldots,f_l](x)}\n
  &=\text{W}\Bigl[
  \frac{\text{W}[f_1,\ldots,f_l,u_1]}{\text{W}[f_1,\ldots,f_l]},\ldots,
  \frac{\text{W}[f_1,\ldots,f_l,u_m]}{\text{W}[f_1,\ldots,f_l]},
  \frac{\text{W}[f_1,\ldots,f_l,v]}{\text{W}[f_1,\ldots,f_l]}\Bigr](x).
\end{align*}
Dividing this equation by \eqref{QM:cor}, we obtain
\begin{equation}
  \frac{\text{W}[f_1,\ldots,f_l,u_1,\ldots,u_m,v](x)}
  {\text{W}[f_1,\ldots,f_l,u_1,\ldots,u_m](x)}
  =\frac{\text{W}\Bigl[
  \frac{\text{W}[f_1,\ldots,f_l,u_1]}{\text{W}[f_1,\ldots,f_l]},\ldots,
  \frac{\text{W}[f_1,\ldots,f_l,u_m]}{\text{W}[f_1,\ldots,f_l]},
  \frac{\text{W}[f_1,\ldots,f_l,v]}{\text{W}[f_1,\ldots,f_l]}\Bigr](x)} 
  {\text{W}\Bigl[
  \frac{\text{W}[f_1,\ldots,f_l,u_1]}{\text{W}[f_1,\ldots,f_l]},\ldots,
  \frac{\text{W}[f_1,\ldots,f_l,u_m]}{\text{W}[f_1,\ldots,f_l]}\Bigr](x)}.
  \label{Wro:id}
\end{equation}
This shows the equality $\phi_{\mathcal{D}\,n}(x)=
\phi_{\mathcal{D}_{\text{v}}\mathcal{D}_{\text{e}}\,n}(x)$ by the following
replacements:
\begin{equation}
  l=M_{\text{v}},\quad m=M_{\text{e}},\quad
  f_j=\tilde{\phi}_{\text{v}_j},\quad u_j=\phi_{e_j},\quad v=\phi_n.
  \label{replace}
\end{equation}

\subsection{Application to idQM}
\label{sec:app_idQM}

Next let us consider idQM.
The virtual states are studied for the exactly solvable systems whose
eigenfunctions are described by the Wilson and Askey-Wilson \cite{os27},
Meixner-Pollaczek and continuous Hahn \cite{idQMcH} polynomials.
The Hamiltonian $\mathcal{H}$ of idQM has the following form:
\begin{equation}
  \mathcal{H}=\sqrt{V(x)}\,e^{\gamma p}\sqrt{V^*(x)}
  +\!\sqrt{V^*(x)}\,e^{-\gamma p}\sqrt{V(x)}-V(x)-V^*(x),
  \label{idQM:H}
\end{equation}
where $x$ is the coordinate and $p$ is the momentum, $p=-i\frac{d}{dx}$, and
$\gamma$ is a nonzero real constant.
The potential function $V(x)$ is an analytic function of $x$ and
the $*$-operation on an analytic function $f(x)=\sum_na_nx^n$
($a_n\in\mathbb{C}$) is defined by $f^*(x)=\sum_na_n^*x^n$, in which
$a_n^*$ is the complex conjugation of $a_n$.
The function $\heihoukon$ is the square root function as a complex function.
The deformed Hamiltonian $\mathcal{H}_{\mathcal{D}}$ \eqref{HD:Scheq} is
given by \cite{gos,os27,idQMcH}
\begin{align}
  &\mathcal{H}_{\mathcal{D}}
  =\sqrt{V_{\mathcal{D}}(x)}\,e^{\gamma p}\sqrt{V_{\mathcal{D}}^*(x)}
  +\!\sqrt{V_{\mathcal{D}}^*(x)}\,e^{-\gamma p}\sqrt{V_{\mathcal{D}}(x)}
  -V_{\mathcal{D}}(x)-V_{\mathcal{D}}^*(x)+\mathcal{E}_{\mu},
  \label{idQM:HD}\\
  &V_{\mathcal{D}}(x)=
  \sqrt{V(x-i\tfrac{M}{2}\gamma)V^*(x-i\tfrac{M+2}{2}\gamma)}\qquad\quad
  \bigl(\,\mu\eqdef\min\{n\,|\,n\in \mathbb{Z}_{\geq 0}
  \backslash\mathcal{D}_{\text{e}}\}\,\bigr)\n
  &\phantom{V_{\mathcal{D}}(x)=}\times
  \frac{\text{W}_{\gamma}[\psi_1,\ldots,\psi_M](x+i\frac{\gamma}{2})}
  {\text{W}_{\gamma}[\psi_1,\ldots,\psi_M](x-i\frac{\gamma}{2})}\,
  \frac{\text{W}_{\gamma}[\psi_1,\ldots,\psi_M,\phi_{\mu}](x-i\gamma)}
  {\text{W}_{\gamma}[\psi_1,\ldots,\psi_M,\phi_{\mu}](x)},
  \label{idQM:VD}
\end{align}
and its eigenfunctions $\phi_{\mathcal{D}\,n}(x)$ are
given by
\begin{align}
  \phi_{\mathcal{D}\,n}(x)&=
  \biggl(\prod_{j=0}^{M-1}V\bigl(x+i(\tfrac{M}{2}-j)\gamma\bigr)
  V^*\bigl(x-i(\tfrac{M}{2}-j)\gamma\bigr)\biggr)^{\frac14}\n
  &\quad\times
  \frac{\text{W}_{\gamma}[\psi_1,\ldots,\psi_M,\phi_n](x)}
  {\sqrt{\text{W}_{\gamma}[\psi_1,\ldots,\psi_M](x-i\frac{\gamma}{2})
  \text{W}_{\gamma}[\psi_1,\ldots,\psi_M](x+i\frac{\gamma}{2})}}
  \ \ (n\in\mathbb{Z}_{\geq 0}\backslash\mathcal{D}_{\text{e}}).
  \label{idQM:phiDn}
\end{align}
On the other hand, the eigenfunctions and the potential function of
$\mathcal{H}_{\mathcal{D}_{\text{v}}}$ \eqref{HDv:Scheq} are given by
\cite{os27,idQMcH}
\begin{align}
  \phi_{\mathcal{D}_{\text{v}}\,n}(x)&=
  \biggl(\prod_{j=0}^{M_{\text{v}}-1}
  V\bigl(x+i(\tfrac{M_{\text{v}}}{2}-j)\gamma\bigr)
  V^*\bigl(x-i(\tfrac{M_{\text{v}}}{2}-j)\gamma\bigr)\biggr)^{\frac14}\n
  &\quad\times
  \frac{\text{W}_{\gamma}[\tilde{\phi}_{\text{v}_1},\ldots,
  \tilde{\phi}_{\text{v}_{M_{\text{v}}}},\phi_n](x)}
  {\sqrt{\text{W}_{\gamma}[\tilde{\phi}_{\text{v}_1},\ldots,
  \tilde{\phi}_{\text{v}_{M_{\text{v}}}}](x-i\frac{\gamma}{2})
  \text{W}_{\gamma}[\tilde{\phi}_{\text{v}_1},\ldots,
  \tilde{\phi}_{\text{v}_{M_{\text{v}}}}](x+i\frac{\gamma}{2})}}
  \ \ (n\in\mathbb{Z}_{\geq 0}),
  \label{idQM:phiDvn}\\[4pt]
  V_{\mathcal{D}_{\text{v}}}(x)&=
  \sqrt{V\bigl(x-i\tfrac{M_{\text{v}}}{2}\gamma\bigr)
  V^*\bigl(x-i\tfrac{M_{\text{v}}+2}{2}\gamma\bigr)}\n
  &\quad\times
  \frac{\text{W}_{\gamma}[\tilde{\phi}_{\text{v}_1},\ldots,
  \tilde{\phi}_{\text{v}_{M_{\text{v}}}}](x+i\frac{\gamma}{2})}
  {\text{W}_{\gamma}[\tilde{\phi}_{\text{v}_1},\ldots,
  \tilde{\phi}_{\text{v}_{M_{\text{v}}}}](x-i\frac{\gamma}{2})}
  \frac{\text{W}_{\gamma}[\tilde{\phi}_{\text{v}_1},\ldots,
  \tilde{\phi}_{\text{v}_{M_{\text{v}}}},\phi_0](x-i\gamma)}
  {\text{W}_{\gamma}[\tilde{\phi}_{\text{v}_1},\ldots,
  \tilde{\phi}_{\text{v}_{M_{\text{v}}}},\phi_0](x)}.
  \label{idQM:VDv}
\end{align}
So the eigenfunctions of
$\mathcal{H}_{\mathcal{D}_{\text{v}}\mathcal{D}_{\text{e}}}$
\eqref{HDvDe:Scheq} are expressed as
\begin{align}
  \phi_{\mathcal{D}_{\text{v}}\mathcal{D}_{\text{e}}\,n}(x)
  &=\biggl(\prod_{j=0}^{M_{\text{e}}-1}
  V_{\mathcal{D}_{\text{v}}}\bigl(x+i(\tfrac{M_{\text{e}}}{2}-j)\gamma\bigr)
  V_{\mathcal{D}_{\text{v}}}^*\bigl(x-i(\tfrac{M_{\text{e}}}{2}-j)\gamma\bigr)
  \biggr)^{\frac14}
  \qquad(n\in\mathbb{Z}_{\geq 0}\backslash\mathcal{D}_{\text{e}})\n
  &\quad\times
  \frac{\text{W}_{\gamma}[\phi_{\mathcal{D}_{\text{v}}\,e_1},\ldots,
  \phi_{\mathcal{D}_{\text{v}}\,e_{M_{\text{e}}}},
  \phi_{\mathcal{D}_{\text{v}}\,n}](x)}
  {\sqrt{\text{W}_{\gamma}[\phi_{\mathcal{D}_{\text{v}}\,e_1},\ldots,
  \phi_{\mathcal{D}_{\text{v}}\,e_{M_{\text{e}}}}](x-i\frac{\gamma}{2})
  \text{W}_{\gamma}[\phi_{\mathcal{D}_{\text{v}}\,e_1},\ldots,
  \phi_{\mathcal{D}_{\text{v}}\,e_{M_{\text{e}}}}](x+i\frac{\gamma}{2})}}.
  \label{idQM:phiDvDen}
\end{align}
We will show that two expressions \eqref{idQM:phiDn} and \eqref{idQM:phiDvDen}
are actually identical by using the Casoratian identity,
Corollary\,\ref{cor:idQM}.

Corollary\,\ref{cor:idQM} is
\begin{align*}
  &\quad\text{W}_{\gamma}[f_1,\ldots,f_l,u_1,\ldots,u_m](x)
  \quad\Bigl(w(x)\eqdef\sqrt{
  \text{W}_{\gamma}[f_1,\ldots,f_l](x-i\tfrac{\gamma}{2})
  \text{W}_{\gamma}[f_1,\ldots,f_l](x+i\tfrac{\gamma}{2})}\ \Bigr)\\
  &=\sqrt{\text{W}_{\gamma}[f_1,\ldots,f_l](x-i\tfrac{m}{2}\gamma)
  \text{W}_{\gamma}[f_1,\ldots,f_l](x+i\tfrac{m}{2}\gamma)}\n
  &\quad\times
  \text{W}_{\gamma}\Bigl[
  \frac{\text{W}_{\gamma}[f_1,\ldots,f_l,u_1]}{w},\ldots,
  \frac{\text{W}_{\gamma}[f_1,\ldots,f_l,u_m]}{w}\Bigr](x),
\end{align*}
and, by the replacements $m\to m+1$ and $u_{m+1}=v$, it becomes
\begin{align*}
  &\quad\text{W}_{\gamma}[f_1,\ldots,f_l,u_1,\ldots,u_m,v](x)\n[2pt]
  &=\sqrt{\text{W}_{\gamma}[f_1,\ldots,f_l](x-i\tfrac{m+1}{2}\gamma)
  \text{W}_{\gamma}[f_1,\ldots,f_l](x+i\tfrac{m+1}{2}\gamma)}\n
  &\quad\times
  \text{W}_{\gamma}\Bigl[
  \frac{\text{W}_{\gamma}[f_1,\ldots,f_l,u_1]}{w},\ldots,
  \frac{\text{W}_{\gamma}[f_1,\ldots,f_l,u_m]}{w},
  \frac{\text{W}_{\gamma}[f_1,\ldots,f_l,v]}{w}\Bigr](x).
\end{align*}
From these two equations, we obtain
\begin{align}
  &\quad\frac{\text{W}_{\gamma}[f_1,\ldots,f_l,u_1,\ldots,u_m,v](x)}
  {\sqrt{\text{W}_{\gamma}[f_1,\ldots,f_l,u_1,\ldots,u_m](x-i\frac{\gamma}{2})
  \text{W}_{\gamma}[f_1,\ldots,f_l,u_1,\ldots,u_m](x+i\frac{\gamma}{2})}}\n
  &=\biggl(\frac{
  \text{W}_{\gamma}[f_1,\ldots,f_l](x-i\frac{m+1}{2}\gamma)
  \text{W}_{\gamma}[f_1,\ldots,f_l](x+i\frac{m+1}{2}\gamma)}
  {\text{W}_{\gamma}[f_1,\ldots,f_l](x-i\frac{m-1}{2}\gamma)
  \text{W}_{\gamma}[f_1,\ldots,f_l](x+i\frac{m-1}{2}\gamma)}
  \biggr)^{\frac14}
  \label{Cas:id}\\
  &\quad\times\frac{\text{W}_{\gamma}\Bigl[
  \frac{\text{W}_{\gamma}[f_1,\ldots,f_l,u_1]}{w},\ldots,
  \frac{\text{W}_{\gamma}[f_1,\ldots,f_l,u_m]}{w},
  \frac{\text{W}_{\gamma}[f_1,\ldots,f_l,v]}{w}\Bigr](x)}
  {\sqrt{\text{W}_{\gamma}\Bigl[
  \frac{\text{W}_{\gamma}[f_1,\ldots,f_l,u_1]}{w},\ldots,
  \frac{\text{W}_{\gamma}[f_1,\ldots,f_l,u_m]}{w}\Bigr](x-i\frac{\gamma}{2})
  \text{W}_{\gamma}\Bigl[
  \frac{\text{W}_{\gamma}[f_1,\ldots,f_l,u_1]}{w},\ldots,
  \frac{\text{W}_{\gamma}[f_1,\ldots,f_l,u_m]}{w}\Bigr](x+i\frac{\gamma}{2})}}.
  \nonumber
\end{align}
In the following, we consider the replacements (identification) \eqref{replace}.
The eigenfunctions \eqref{idQM:phiDvn} of
$\mathcal{H}_{\mathcal{D}_{\text{v}}}$ are expressed as
\begin{align}
  &\phi_{\mathcal{D}_{\text{v}}\,n}(x)=G(x)
  \frac{\text{W}_{\gamma}[f_1,\ldots,f_l,v](x)}{w(x)},\quad
  \phi_{\mathcal{D}_{\text{v}}\,e_j}(x)=G(x)
  \frac{\text{W}_{\gamma}[f_1,\ldots,f_l,u_j](x)}{w(x)},\\
  &G(x)\eqdef\Bigl(\prod_{j=0}^{l-1}
  V\bigl(x+i(\tfrac{l}{2}-j)\gamma\bigr)
  V^*\bigl(x-i(\tfrac{l}{2}-j)\gamma\bigr)\Bigr)^{\frac14}.
\end{align}
By Proposition\,\ref{prop:idQM:1}, we have
\begin{align}
  &\quad\frac{\text{W}_{\gamma}[\phi_{\mathcal{D}_{\text{v}}\,e_1},\ldots,
  \phi_{\mathcal{D}_{\text{v}}\,e_{M_{\text{e}}}},
  \phi_{\mathcal{D}_{\text{v}}\,n}](x)}
  {\sqrt{\text{W}_{\gamma}[\phi_{\mathcal{D}_{\text{v}}\,e_1},\ldots,
  \phi_{\mathcal{D}_{\text{v}}\,e_{M_{\text{e}}}}](x-i\frac{\gamma}{2})
  \text{W}_{\gamma}[\phi_{\mathcal{D}_{\text{v}}\,e_1},\ldots,
  \phi_{\mathcal{D}_{\text{v}}\,e_{M_{\text{e}}}}](x+i\frac{\gamma}{2})}}\n
  &=\frac{\text{W}_{\gamma}\Bigl[
  \frac{\text{W}_{\gamma}[f_1,\ldots,f_l,u_1]}{w},\ldots,
  \frac{\text{W}_{\gamma}[f_1,\ldots,f_l,u_m]}{w},
  \frac{\text{W}_{\gamma}[f_1,\ldots,f_l,v]}{w}\Bigr](x)}
  {\sqrt{\text{W}_{\gamma}\Bigl[
  \frac{\text{W}_{\gamma}[f_1,\ldots,f_l,u_1]}{w},\ldots,
  \frac{\text{W}_{\gamma}[f_1,\ldots,f_l,u_m]}{w}\Bigr](x-i\frac{\gamma}{2})
  \text{W}_{\gamma}\Bigl[
  \frac{\text{W}_{\gamma}[f_1,\ldots,f_l,u_1]}{w},\ldots,
  \frac{\text{W}_{\gamma}[f_1,\ldots,f_l,u_m]}{w}\Bigr](x+i\frac{\gamma}{2})}}
  \n
  &\quad\times\frac{\prod_{j=1}^{m+1}G(x^{(m+1)}_j)}
  {\sqrt{\prod_{j=1}^{m}G(x^{(m)}_j-i\frac{\gamma}{2})
  G(x^{(m)}_j+i\frac{\gamma}{2})}},
  \label{W/sqrtWW}
\end{align}
and a short calculation shows
\begin{align}
  \frac{\prod_{j=1}^{m+1}G(x^{(m+1)}_j)}
  {\sqrt{\prod_{j=1}^{m}G(x^{(m)}_j-i\frac{\gamma}{2})
  G(x^{(m)}_j+i\frac{\gamma}{2})}}
  &=\biggl(\prod_{j=0}^{l-1}
  V\bigl(x+i(\tfrac{l+m}{2}-j)\gamma\bigr)
  V^*\bigl(x-i(\tfrac{l+m}{2}-j)\gamma\bigr)
  \label{G/sqrtGG}\\
  &\qquad\times
  \prod_{j=m}^{l+m-1}
  V\bigl(x+i(\tfrac{l+m}{2}-j)\gamma\bigr)
  V^*\bigl(x-i(\tfrac{l+m}{2}-j)\gamma\bigr)
  \biggr)^{\frac18}.
  \nonumber
\end{align}
For the potential function $V_{\mathcal{D}_{\text{v}}}(x)$ \eqref{idQM:VDv},
a short calculation shows
\begin{align}
  &\quad\prod_{j=0}^{m-1}
  V_{\mathcal{D}_{\text{v}}}\bigl(x+i(\tfrac{m}{2}-j)\gamma\bigr)
  V^*_{\mathcal{D}_{\text{v}}}\bigl(x-i(\tfrac{m}{2}-j)\gamma\bigr)\n
  &=\biggl(\prod_{j=0}^{m-1}
  V\bigl(x+i(\tfrac{l+m}{2}-j)\gamma\bigr)
  V^*\bigl(x-i(\tfrac{l+m}{2}-j)\gamma\bigr)\n
  &\qquad\times
  \prod_{j=l}^{l+m-1}
  V\bigl(x+i(\tfrac{l+m}{2}-j)\gamma\bigr)
  V^*\bigl(x-i(\tfrac{l+m}{2}-j)\gamma\bigr)
  \biggr)^{\frac12}\n
  &\quad\times\frac{
  \text{W}_{\gamma}[f_1,\ldots,f_l](x-i\frac{m+1}{2}\gamma)
  \text{W}_{\gamma}[f_1,\ldots,f_l](x+i\frac{m+1}{2}\gamma)}
  {\text{W}_{\gamma}[f_1,\ldots,f_l](x-i\frac{m-1}{2}\gamma)
  \text{W}_{\gamma}[f_1,\ldots,f_l](x+i\frac{m-1}{2}\gamma)}.
  \label{VDvVDvs}
\end{align}
{}From \eqref{VDvVDvs}, \eqref{W/sqrtWW}--\eqref{G/sqrtGG} and
\eqref{Cas:id}, we obtain
\begin{align}
  \eqref{idQM:phiDvDen}&=
  \biggl(\prod_{j=0}^{l+m-1}V\bigl(x+i(\tfrac{l+m}{2}-j)\gamma\bigr)
  V^*\bigl(x-i(\tfrac{l+m}{2}-j)\gamma\bigr)\biggr)^{\frac14}\n
  &\quad\times
  \frac{\text{W}_{\gamma}[f_1,\ldots,f_l,u_1,\ldots,u_m,v](x)}
  {\sqrt{\text{W}_{\gamma}[f_1,\ldots,f_l,u_1,\ldots,u_m](x-i\frac{\gamma}{2})
  \text{W}_{\gamma}[f_1,\ldots,f_l,u_1,\ldots,u_m](x+i\frac{\gamma}{2})}}\\
  &=\eqref{idQM:phiDn},
  \nonumber
\end{align}
namely the equality $\phi_{\mathcal{D}\,n}(x)=
\phi_{\mathcal{D}_{\text{v}}\mathcal{D}_{\text{e}}\,n}(x)$.

\subsection{Application to rdQM}
\label{sec:app_rdQM}

Next let us consider rdQM.
The virtual states are studied for the exactly solvable systems whose
eigenfunctions are described by the Racah and $q$-Racah \cite{os26},
Meixner and little $q$-Jacobi (Laguerre) \cite{os35} polynomials.
The Hamiltonian of rdQM, $\mathcal{H}=(\mathcal{H}_{x,y})$, is a tri-diagonal
real symmetric (Jacobi) matrix and its rows and columns are indexed by
integers $x$ and $y$, which take values in $\{0,1,\ldots,x_{\text{max}}\}$
(finite) or $\mathbb{Z}_{\geq 0}$ (semi-infinite) or $\mathbb{Z}$
(full infinite),
\begin{equation}
  \mathcal{H}_{x,y}=
  -\sqrt{B(x)D(x+1)}\,\delta_{x+1,y}-\sqrt{B(x-1)D(x)}\,\delta_{x-1,y}
  +\bigl(B(x)+D(x)\bigr)\delta_{x,y}.
  \label{rdQM:Hxy}
\end{equation}
There exist finite and semi-infinite rdQM systems with virtual states
\cite{os26,os35}, but we do not know full infinite rdQM systems with virtual
states.
In the following, for simplicity of presentation, we consider semi-infinite
systems only (For finite systems, some modification is needed).
The potential functions $B(x)$ and $D(x)$ are real and positive but vanish
at the boundary: $B(x)>0$ ($n\in\mathbb{Z}_{\geq 0}$),
$D(x)>0$ ($n\in\mathbb{Z}_{\geq 1}$) and $D(0)=0$.
The function $\heihoukon$ is the square root function as a real function.
We take the normalization of $\phi_n(x)$ \eqref{H:Scheq} and
$\tilde{\phi}_{\text{v}}(x)$ \eqref{vs:Scheq} of the original system as
$\phi_n(0)=\tilde{\phi}_{\text{v}}(0)=1$.
For simplicity in notation, we write the matrix $\mathcal{H}$ as follows:
\begin{equation}
  \mathcal{H}=-\sqrt{B(x)}\,e^{\partial}\sqrt{D(x)}
  -\!\sqrt{D(x)}\,e^{-\partial}\sqrt{B(x)}+B(x)+D(x),
  \label{rdQM:H}
\end{equation}
where matrices $e^{\pm\partial}$ are
$(e^{\pm\partial})_{x,y}\eqdef\delta_{x\pm 1,y}$ and
the unit matrix $\bm{1}=(\delta_{x,y})$ is suppressed.
The notation $f(x)Ag(x)$, where $f(x)$ and $g(x)$ are functions of $x$ and
$A$ is a matrix $A=(A_{x,y})$, stands for a matrix whose $(x,y)$-element
is $f(x)A_{x,y}g(y)$.
Note that the matrices $e^{\partial}$ and $e^{-\partial}$ are not inverse
to each other: $e^{\partial}e^{-\partial}=\bm{1}$ but
$e^{-\partial}e^{\partial}\neq\bm{1}$.
This Hamiltonian can be expressed in a factorized form:
\begin{equation}
  \mathcal{H}=\mathcal{A}^{\dagger}\mathcal{A},\quad
  \mathcal{A}\eqdef\sqrt{B(x)}-e^{\partial}\sqrt{D(x)},\quad
  \mathcal{A}^{\dagger}=\sqrt{B(x)}-\sqrt{D(x)}\,e^{-\partial}.
  \label{factor}
\end{equation}
The deformed Hamiltonian $\mathcal{H}_{\mathcal{D}}$ \eqref{HD:Scheq}
is given by \cite{os22,os26,os35}
\begin{align}
  &\mathcal{H}_{\mathcal{D}}
  =-\sqrt{B_{\mathcal{D}}(x)}\,e^{\partial}\sqrt{D_{\mathcal{D}}(x)}
  -\!\sqrt{D_{\mathcal{D}}(x)}\,e^{-\partial}\sqrt{B_{\mathcal{D}}(x)}
  +B_{\mathcal{D}}(x)+D_{\mathcal{D}}(x)+\mathcal{E}_{\mu}
  \label{rdQM:HD}\\
  &\phantom{\mathcal{H}_{\mathcal{D}}}
  =\mathcal{A}_{\mathcal{D}}^{\dagger}\mathcal{A}_{\mathcal{D}}
  +\mathcal{E}_{\mu},\qquad\quad
  \bigl(\,\mu\eqdef\min\{n\,|\,n\in \mathbb{Z}_{\geq 0}
  \backslash\mathcal{D}_{\text{e}}\}\,\bigr),
  \nonumber
\end{align}
where the potential functions $B_{\mathcal{D}}(x)$ and $D_{\mathcal{D}}(x)$ are
\begin{align}
  B_{\mathcal{D}}(x)&=
  \sqrt{B(x+M)D(x+M+1)}\,
  \frac{\text{W}_{\text{C}}[\psi_1,\ldots,\psi_M](x)}
  {\text{W}_{\text{C}}[\psi_1,\ldots,\psi_M](x+1)}\,
  \frac{\text{W}_{\text{C}}[\psi_1,\ldots,\psi_M,\phi_{\mu}](x+1)}
  {\text{W}_{\text{C}}[\psi_1,\ldots,\psi_M,\phi_{\mu}](x)},\n
  D_{\mathcal{D}}(x)&=
  \sqrt{B(x-1)D(x)}\,
  \frac{\text{W}_{\text{C}}[\psi_1,\ldots,\psi_M](x+1)}
  {\text{W}_{\text{C}}[\psi_1,\ldots,\psi_M](x)}\,
  \frac{\text{W}_{\text{C}}[\psi_1,\ldots,\psi_M,\phi_{\mu}](x-1)}
  {\text{W}_{\text{C}}[\psi_1,\ldots,\psi_M,\phi_{\mu}](x)}.
  \label{rdQM:BDD}
\end{align}
Its eigenfunctions $\phi_{\mathcal{D}\,n}(x)$ are given by
\begin{align}
  \phi_{\mathcal{D}\,n}(x)&=(-1)^M\epsilon_{\mathcal{D}}
  \biggl(\prod_{j=1}^MB(x+j-1)D(x+j)\biggr)^{\frac14}\n
  &\quad\times
  \frac{\text{W}_{\text{C}}[\psi_1,\ldots,\psi_M,\phi_n](x)}
  {\sqrt{\text{W}_{\text{C}}[\psi_1,\ldots,\psi_M](x)
  \text{W}_{\text{C}}[\psi_1,\ldots,\psi_M](x+1)}},
  \label{rdQM:phiDn}
\end{align}
where the sign factor $\epsilon_{\mathcal{D}}$ is defined by
\begin{equation}
  \epsilon_{\mathcal{D}}=\epsilon_{d_1\ldots d_M}
  \eqdef\!\!\prod_{1\leq i<j\leq M}\!\!\!\!\text{sgn}\,
  (\mathcal{E}_{\psi_i}-\mathcal{E}_{\psi_j}),
\end{equation}
(for $M=0,1$, we set $\epsilon_{\mathcal{D}}=1$. $\mathcal{D}$ is regarded
as an ordered set.).
Here $\mathcal{E}_{\psi_j}$ is
$\mathcal{E}_{\psi_j}=\tilde{\mathcal{E}}_{\text{v}_k}$ for $d_j=\text{v}_k$
and $\mathcal{E}_{\psi_j}=\mathcal{E}_{e_k}$ for $d_j=e_k$.
This sign factor $\epsilon_{\mathcal{D}}$ was written as
$(-1)^M\mathcal{S}_{d_1\ldots d_M}$ in \cite{os35}, but we missed it
in \cite{os22,os26}.
The sign factor $\epsilon_{\mathcal{D}}$ is important for Darboux
transformations, but not as an eigenfunction.

Before we go any further, let us mention the square root function and
the sign of $\text{W}_{\text{C}}[\psi_1,\ldots,\psi_M](x)$.
If the Krein-Adler condition \eqref{KAcond} is satisfied and the range of
parameters is chosen appropriately, we have the following two facts
(conjectures for $\mathcal{D}_{\text{e}}\neq\emptyset$ case, which
are supported by numerical calculation).
(\romannumeral1) : The potential functions $B_{\mathcal{D}}(x)$ and
$D_{\mathcal{D}}(x)$ are real and positive (except for $D_{\mathcal{D}}(0)=0$).
(\romannumeral2) : The function
$\text{W}_{\text{C}}[\psi_1,\ldots,\psi_M](x)$ has a definite sign
$\epsilon_{d_1\ldots d_M}$, namely
$\text{sgn}\,\text{W}_{\text{C}}[\psi_1,\ldots,\psi_M](x)
=\epsilon_{d_1\ldots d_M}$ ($x\in\mathbb{Z}_{\geq 0}$).
The fact (\romannumeral1) means that $\mathcal{H}_{\mathcal{D}}$
\eqref{rdQM:HD} is well-defined and hermitian, and (\romannumeral2) implies
that $\phi_{\mathcal{D}\,n}(x)$ \eqref{rdQM:phiDn} is real, because
$\text{W}_{\text{C}}[\psi_1,\ldots,\psi_M](x)
\text{W}_{\text{C}}[\psi_1,\ldots,\psi_M](x+1)$
in the square root is positive.
However, in the intermediate steps of the multi-step Darboux transformations
with $\mathcal{D}_{\text{e}}\neq\emptyset$, the Krein-Adler condition
\eqref{KAcond} may not be satisfied. This means that the function
$\text{W}_{\text{C}}[\psi_1,\ldots,\psi_{M'}](x)$ ($M'<M$) may not have
a definite sign.
If so, the argument of the square root in \eqref{rdQM:phiDn} (with $M\to M'$)
becomes negative, and the potential functions $B_{\mathcal{D}}(x)$ and
$D_{\mathcal{D}}(x)$ (with $M\to M'$) also become negative.
Since we regard $\heihoukon$ as a real function, its argument should be real
and non-negative, and its value is also real and non-negative.
We remark that the final result \eqref{HD:Scheq}, which is obtained by the
$M$-step Darboux transformations with $\mathcal{D}$ satisfying the Krein-Adler
condition \eqref{KAcond}, is correct, because the calculation of the Darboux
transformation is purely algebraic.
Since the argument of $\heihoukon$ may be negative in the intermediate steps,
we have to specify how to treat $\sqrt{f(x)}$ for the function $f(x)$ that
does not have a definite sign.
We missed pointing out this remark in \cite{os22}.

We adopt the following rule for $\sqrt{f(x)}$.
If it is not necessary, the value of $\sqrt{f(x)}$ is not evaluated and is
left as it is.
By using the property $\sqrt{a}\,\sqrt{b}=\sqrt{ab}\,$, the calculation is
continued as follows:
$\sqrt{f(x)}/\sqrt{f(x)}=\sqrt{f(x)/f(x)}=\sqrt{1}=1$ and
$\sqrt{f(x)}\,\sqrt{f(x)}=\sqrt{f(x)^2}=\text{sgn}\,f(0)\cdot f(x)$.
We remark that this rule gives correct results for the function with a
definite sign.
Let us illustrate this rule by the calculation on the sign factor
$\epsilon_{\mathcal{D}}$.
We assume that the virtual state energy $\tilde{\mathcal{E}}_{\text{v}}$
\eqref{vs:Scheq} is a monotonically increasing or decreasing function of
$\text{v}$, which is possible by choosing the range of parameters
appropriately.
We assume $\text{sgn}\,\text{W}_{\text{C}}[\psi_1,\ldots,\psi_M](x)
=\epsilon_{\mathcal{D}}$ for $x=0,1$
(for an appropriate range of the parameters)
even if the Krein-Adler condition \eqref{KAcond} is not satisfied.
This assumption can be verified by numerical calculation.
In the intermediate steps of the Darboux transformations, the deformed
Hamiltonian $\mathcal{H}_{d_1\ldots d_s}$, which may be singular, is
\cite{os22,os26,os35}
\begin{align}
  &\mathcal{H}_{d_1\ldots d_s}
  =\hat{\mathcal{A}}_{d_1\dots d_s}\hat{\mathcal{A}}^{\dagger}_{d_1\dots d_s}
  +\mathcal{E}_{\psi_s},\\
  &\hat{\mathcal{A}}_{d_1\ldots d_s}
  =\sqrt{\hat{B}_{d_1\dots d_s}(x)}-e^{\partial}
  \sqrt{\hat{D}_{d_1\dots d_s}(x)},\quad
  \hat{\mathcal{A}}^{\dagger}_{d_1\ldots d_s}
  =\sqrt{\hat{B}_{d_1\dots d_s}(x)}
  -\sqrt{\hat{D}_{d_1\dots d_s}(x)}\,e^{-\partial},
\end{align}
where the potential functions $\hat{B}_{d_1\dots d_s}(x)$ and
$\hat{D}_{d_1\dots d_s}(x)$ are
\begin{align}
  &\hat{B}_{d_1\dots d_s}(x)=
  \sqrt{B(x+s-1)D(x+s)}\,
  \frac{\text{W}_{\text{C}}[\psi_1,\ldots,\psi_{s-1}](x)}
  {\text{W}_{\text{C}}[\psi_1,\ldots,\psi_{s-1}](x+1)}\,
  \frac{\text{W}_{\text{C}}[\psi_1,\ldots,\psi_s](x+1)}
  {\text{W}_{\text{C}}[\psi_1,\ldots,\psi_s](x)},\n
  &\hat{D}_{d_1\dots d_s}(x)=
  \sqrt{B(x-1)D(x)}\,
  \frac{\text{W}_{\text{C}}[\psi_1,\ldots,\psi_{s-1}](x+1)}
  {\text{W}_{\text{C}}[\psi_1,\ldots,\psi_{s-1}](x)}\,
  \frac{\text{W}_{\text{C}}[\psi_1,\ldots,\psi_s](x-1)}
  {\text{W}_{\text{C}}[\psi_1,\ldots,\psi_s](x)}.
\end{align}
Its ``eigenfunctions'' $\phi_{d_1\ldots d_s\,n}(x)$ are
\begin{align}
  &\quad\phi_{d_1\dots d_s\,n}(x)
  \eqdef\hat{\mathcal{A}}_{d_1\ldots d_s}\phi_{d_1\ldots d_{s-1}\,n}\\
  &=(-1)^s\epsilon_{d_1\dots d_s}
  \biggl(\prod_{j=1}^sB(x+j-1)D(x+j)\biggr)^{\frac14}
  \frac{\text{W}_{\text{C}}[\psi_1,\ldots,\psi_s,\phi_n](x)}
  {\sqrt{\text{W}_{\text{C}}[\psi_1,\ldots,\psi_s](x)
  \text{W}_{\text{C}}[\psi_1,\ldots,\psi_s](x+1)}}.
  \nonumber
\end{align}
By calculation with careful treatment of the square root, the next step
``eigenfunction'' $\phi_{d_1\ldots d_{s+1}\,n}(x)$ becomes
\begin{align}
  &\quad\phi_{d_1\ldots d_{s+1}\,n}(x)
  =\hat{\mathcal{A}}_{d_1\ldots d_{s+1}}\phi_{d_1\ldots d_s\,n}(x)\n
  &=\sqrt{\sqrt{B(x+s)D(x+s+1)}\,
  \frac{\text{W}_{\text{C}}[\psi_1,\ldots,\psi_s](x)}
  {\text{W}_{\text{C}}[\psi_1,\ldots,\psi_s](x+1)}
  \frac{\text{W}_{\text{C}}[\psi_1,\ldots,\psi_{s+1}](x+1)}
  {\text{W}_{\text{C}}[\psi_1,\ldots,\psi_{s+1}](x)}}\n
  &\quad\times(-1)^s\epsilon_{d_1\ldots d_s}
  \biggl(\prod_{j=1}^sB(x+j-1)D(x+j)\biggr)^{\frac14}
  \frac{\text{W}_{\text{C}}[\psi_1,\ldots,\psi_s,\phi_n](x)}
  {\sqrt{\text{W}_{\text{C}}[\psi_1,\ldots,\psi_s](x)
  \text{W}_{\text{C}}[\psi_1,\ldots,\psi_s](x+1)}}\n
  &\quad-\sqrt{\sqrt{B(x)D(x+1)}\,
  \frac{\text{W}_{\text{C}}[\psi_1,\ldots,\psi_s](x+2)}
  {\text{W}_{\text{C}}[\psi_1,\ldots,\psi_s](x+1)}
  \frac{\text{W}_{\text{C}}[\psi_1,\ldots,\psi_{s+1}](x)}
  {\text{W}_{\text{C}}[\psi_1,\ldots,\psi_{s+1}](x+1)}}\n
  &\quad\times(-1)^s\epsilon_{d_1\ldots d_s}
  \biggl(\prod_{j=1}^sB(x+j)D(x+j+1)\biggr)^{\frac14}
  \frac{\text{W}_{\text{C}}[\psi_1,\ldots,\psi_s,\phi_n](x+1)}
  {\sqrt{\text{W}_{\text{C}}[\psi_1,\ldots,\psi_s](x+1)
  \text{W}_{\text{C}}[\psi_1,\ldots,\psi_s](x+2)}}\n
  &\stackrel{(\text{\romannumeral1})}{=}(-1)^{s+1}\epsilon_{d_1\ldots d_s}
  \biggl(\prod_{j=1}^{s+1}B(x+j-1)D(x+j)\biggr)^{\frac14}
  \frac{1}{\sqrt{\text{W}_{\text{C}}[\psi_1,\ldots,\psi_{s+1}](x)
  \text{W}_{\text{C}}[\psi_1,\ldots,\psi_{s+1}](x+1)}}\n
  &\quad\times
  \frac{1}{\sqrt{\text{W}_{\text{C}}[\psi_1,\ldots,\psi_s](x+1)^2}}
  \biggl(\sqrt{\text{W}_{\text{C}}[\psi_1,\ldots,\psi_{s+1}](x)^2}\,
  \text{W}_{\text{C}}[\psi_1,\ldots,\psi_s,\phi_n](x+1)\n
  &\phantom{\quad\times
  \frac{1}{\sqrt{\text{W}_{\text{C}}[\psi_1,\ldots,\psi_s](x+1)^2}}
  \biggl(}
  -\sqrt{\text{W}_{\text{C}}[\psi_1,\ldots,\psi_{s+1}](x+1)^2}\,
  \text{W}_{\text{C}}[\psi_1,\ldots,\psi_s,\phi_n](x)\biggr)\n
  &\stackrel{(\text{\romannumeral2})}{=}(-1)^{s+1}\epsilon_{d_1\dots d_s}
  \biggl(\prod_{j=1}^{s+1}B(x+j-1)D(x+j)\biggr)^{\frac14}
  \frac{1}{\sqrt{\text{W}_{\text{C}}[\psi_1,\ldots,\psi_{s+1}](x)
  \text{W}_{\text{C}}[\psi_1,\ldots,\psi_{s+1}](x+1)}}\n
  &\quad\times
  \frac{\epsilon_{d_1\ldots d_{s+1}}}{\epsilon_{d_1\ldots d_s}}
  \text{W}_{\text{C}}[\psi_1,\ldots,\psi_{s+1},\phi_n](x)\n
  &=(-1)^{s+1}\epsilon_{d_1\dots d_{s+1}}
  \biggl(\prod_{j=1}^{s+1}B(x+j-1)D(x+j)\biggr)^{\frac14}\n
  &\quad\times
  \frac{\text{W}_{\text{C}}[\psi_1,\ldots,\psi_{s+1},\phi_n](x)}
  {\sqrt{\text{W}_{\text{C}}[\psi_1,\ldots,\psi_{s+1}](x)
  \text{W}_{\text{C}}[\psi_1,\ldots,\psi_{s+1}](x+1)}},
\end{align}
where we have used (\romannumeral1): $\sqrt{a}\,\sqrt{b}=\sqrt{ab}$,
(\romannumeral2): the rule $\sqrt{f(x)^2}=\text{sgn}\,f(0)\cdot f(x)$ and
the Casoratian identity \eqref{rdQM:Cformula}.
This calculation establishes the sign factor $\epsilon_{\mathcal{D}}$ in
\eqref{rdQM:phiDn}.

Let's return to the main topic of this subsection.
The eigenfunctions $\phi_{\mathcal{D}\,n}(x)$ \eqref{HD:Scheq} are given by
\eqref{rdQM:phiDn}.
On the other hand, the eigenfunctions of
$\mathcal{H}_{\mathcal{D}_{\text{v}}}$ \eqref{HDv:Scheq} are given by
\cite{os26,os35}
\begin{align}
  \phi_{\mathcal{D}_{\text{v}}\,n}(x)&=
  (-1)^{M_{\text{v}}}\epsilon_{\mathcal{D}_{\text{v}}}
  \biggl(\prod_{j=1}^{M_{\text{v}}}
  B(x+j-1)D(x+j)\biggr)^{\frac14}\n
  &\quad\times
  \frac{\text{W}_{\text{C}}[\tilde{\phi}_{\text{v}_1},\ldots,
  \tilde{\phi}_{\text{v}_{M_{\text{v}}}},\phi_n](x)}
  {\sqrt{\text{W}_{\text{C}}[\tilde{\phi}_{\text{v}_1},\ldots,
  \tilde{\phi}_{\text{v}_{M_{\text{v}}}}](x)
  \text{W}_{\text{C}}[\tilde{\phi}_{\text{v}_1},\ldots,
  \tilde{\phi}_{\text{v}_{M_{\text{v}}}}](x+1)}}
  \ \ (n\in\mathbb{Z}_{\geq 0}),
  \label{rdQM:phiDvn}
\end{align}
and the potential functions of $\mathcal{H}_{\mathcal{D}_{\text{v}}}$ are
\begin{align}
  B_{\mathcal{D}_{\text{v}}}(x)&=
  \sqrt{B(x+M_{\text{v}})D(x+M_{\text{v}}+1)}\n
  &\quad\times
  \frac{\text{W}_{\text{C}}[\tilde{\phi}_{\text{v}_1},\ldots,
  \tilde{\phi}_{\text{v}_{M_{\text{v}}}}](x)}
  {\text{W}_{\text{C}}[\tilde{\phi}_{\text{v}_1},\ldots,
  \tilde{\phi}_{\text{v}_{M_{\text{v}}}}](x+1)}
  \frac{\text{W}_{\text{C}}[\tilde{\phi}_{\text{v}_1},\ldots,
  \tilde{\phi}_{\text{v}_{M_{\text{v}}}},\phi_0](x+1)}
  {\text{W}_{\text{C}}[\tilde{\phi}_{\text{v}_1},\ldots,
  \tilde{\phi}_{\text{v}_{M_{\text{v}}}},\phi_0](x)},\n
%
  D_{\mathcal{D}_{\text{v}}}(x)&=
  \sqrt{B(x-1)D(x)}\n
  &\quad\times
  \frac{\text{W}_{\text{C}}[\tilde{\phi}_{\text{v}_1},\ldots,
  \tilde{\phi}_{\text{v}_{M_{\text{v}}}}](x+1)}
  {\text{W}_{\text{C}}[\tilde{\phi}_{\text{v}_1},\ldots,
  \tilde{\phi}_{\text{v}_{M_{\text{v}}}}](x)}
  \frac{\text{W}_{\text{C}}[\tilde{\phi}_{\text{v}_1},\ldots,
  \tilde{\phi}_{\text{v}_{M_{\text{v}}}},\phi_0](x-1)}
  {\text{W}_{\text{C}}[\tilde{\phi}_{\text{v}_1},\ldots,
  \tilde{\phi}_{\text{v}_{M_{\text{v}}}},\phi_0](x)}.
  \label{rdQM:DDv}
\end{align}
So the eigenfunctions of
$\mathcal{H}_{\mathcal{D}_{\text{v}}\mathcal{D}_{\text{e}}}$
\eqref{HDvDe:Scheq} are expressed as
\begin{align}
  \phi_{\mathcal{D}_{\text{v}}\mathcal{D}_{\text{e}}\,n}(x)
  &=(-1)^{M_{\text{e}}}\epsilon_{\mathcal{D}_{\text{e}}}
  \biggl(\prod_{j=1}^{M_{\text{e}}}
  B_{\mathcal{D}_{\text{v}}}(x+j-1)
  D_{\mathcal{D}_{\text{v}}}(x+j)
  \biggr)^{\frac14}
  \qquad(n\in\mathbb{Z}_{\geq 0}\backslash\mathcal{D}_{\text{e}})\n
  &\quad\times
  \frac{\text{W}_{\text{C}}[\phi_{\mathcal{D}_{\text{v}}\,e_1},\ldots,
  \phi_{\mathcal{D}_{\text{v}}\,e_{M_{\text{e}}}},
  \phi_{\mathcal{D}_{\text{v}}\,n}](x)}
  {\sqrt{\text{W}_{\text{C}}[\phi_{\mathcal{D}_{\text{v}}\,e_1},\ldots,
  \phi_{\mathcal{D}_{\text{v}}\,e_{M_{\text{e}}}}](x)
  \text{W}_{\text{C}}[\phi_{\mathcal{D}_{\text{v}}\,e_1},\ldots,
  \phi_{\mathcal{D}_{\text{v}}\,e_{M_{\text{e}}}}](x+1)}}.
  \label{rdQM:phiDvDen}
\end{align}
We will show that two expressions \eqref{rdQM:phiDn} and \eqref{rdQM:phiDvDen}
are actually identical by using the Casoratian identity,
Corollary\,\ref{cor:rdQM}.

As noted in the Remark below Corollary\,\ref{cor:rdQM},
Corollary\,\ref{cor:rdQM} has been shown for
$\text{W}_{\text{C}}[f_1,\ldots,f_l](x)$ $>0$.
If $\text{W}_{\text{C}}[f_1,\ldots,f_l](x)$ is a definite sign function
with sign $\epsilon$, Corollary\,\ref{cor:rdQM} becomes
\begin{align*}
  &\quad\text{W}_{\text{C}}[f_1,\ldots,f_l,u_1,\ldots,u_m](x)
  \quad\Bigl(w(x)\eqdef\sqrt{
  \text{W}_{\text{C}}[f_1,\ldots,f_l](x)
  \text{W}_{\text{C}}[f_1,\ldots,f_l](x+1)}\ \Bigr)\\
  &=\epsilon^{m-1}\sqrt{\text{W}_{\text{C}}[f_1,\ldots,f_l](x)
  \text{W}_{\text{C}}[f_1,\ldots,f_l](x+m)}\n
  &\quad\times
  \text{W}_{\text{C}}\Bigl[
  \frac{\text{W}_{\text{C}}[f_1,\ldots,f_l,u_1]}{w},\ldots,
  \frac{\text{W}_{\text{C}}[f_1,\ldots,f_l,u_m]}{w}\Bigr](x),
\end{align*}
and, by the replacements $m\to m+1$ and $u_{m+1}=v$, it becomes
\begin{align*}
  &\quad\text{W}_{\text{C}}[f_1,\ldots,f_l,u_1,\ldots,u_m,v](x)\n[2pt]
  &=\epsilon^m\sqrt{\text{W}_{\text{C}}[f_1,\ldots,f_l](x)
  \text{W}_{\text{C}}[f_1,\ldots,f_l](x+m+1)}\n
  &\quad\times
  \text{W}_{\text{C}}\Bigl[
  \frac{\text{W}_{\text{C}}[f_1,\ldots,f_l,u_1]}{w},\ldots,
  \frac{\text{W}_{\text{C}}[f_1,\ldots,f_l,u_m]}{w},
  \frac{\text{W}_{\text{C}}[f_1,\ldots,f_l,v]}{w}\Bigr](x).
\end{align*}
From these two equations, we obtain
\begin{align}
  &\quad\frac{\text{W}_{\text{C}}[f_1,\ldots,f_l,u_1,\ldots,u_m,v](x)}
  {\sqrt{\text{W}_{\text{C}}[f_1,\ldots,f_l,u_1,\ldots,u_m](x)
  \text{W}_{\text{C}}[f_1,\ldots,f_l,u_1,\ldots,u_m](x+1)}}\n
  &=\epsilon^m\biggl(\frac{
  \text{W}_{\text{C}}[f_1,\ldots,f_l](x)
  \text{W}_{\text{C}}[f_1,\ldots,f_l](x+m+1)}
  {\text{W}_{\text{C}}[f_1,\ldots,f_l](x+1)
  \text{W}_{\text{C}}[f_1,\ldots,f_l](x+m)}
  \biggr)^{\frac14}
  \label{rdQM:Cas:id}\\
  &\quad\times\frac{\text{W}_{\text{C}}\Bigl[
  \frac{\text{W}_{\text{C}}[f_1,\ldots,f_l,u_1]}{w},\ldots,
  \frac{\text{W}_{\text{C}}[f_1,\ldots,f_l,u_m]}{w},
  \frac{\text{W}_{\text{C}}[f_1,\ldots,f_l,v]}{w}\Bigr](x)}
  {\sqrt{\text{W}_{\text{C}}\Bigl[
  \frac{\text{W}_{\text{C}}[f_1,\ldots,f_l,u_1]}{w},\ldots,
  \frac{\text{W}_{\text{C}}[f_1,\ldots,f_l,u_m]}{w}\Bigr](x)
  \text{W}_{\text{C}}\Bigl[
  \frac{\text{W}_{\text{C}}[f_1,\ldots,f_l,u_1]}{w},\ldots,
  \frac{\text{W}_{\text{C}}[f_1,\ldots,f_l,u_m]}{w}\Bigr](x+1)}}.
  \nonumber
\end{align}
In the following, we consider the replacements (identification) \eqref{replace}.
The sign factor $\epsilon$ in \eqref{rdQM:Cas:id} becomes
$\epsilon=\epsilon_{\mathcal{D}_{\text{v}}}$.
The eigenfunctions \eqref{rdQM:phiDvn} of
$\mathcal{H}_{\mathcal{D}_{\text{v}}}$ are expressed as
\begin{align}
  &\phi_{\mathcal{D}_{\text{v}}\,n}(x)=G(x)
  \frac{\text{W}_{\text{C}}[f_1,\ldots,f_l,v](x)}{w(x)},\quad
  \phi_{\mathcal{D}_{\text{v}}\,e_j}(x)=G(x)
  \frac{\text{W}_{\text{C}}[f_1,\ldots,f_l,u_j](x)}{w(x)},\\
  &G(x)\eqdef(-1)^l\epsilon_{D_{\text{v}}}
  \Bigl(\prod_{j=1}^lB(x+j-1)D(x+j)\Bigr)^{\frac14}.
\end{align}
By Proposition\,\ref{prop:rdQM:1}, we have
\begin{align}
  &\quad\frac{\text{W}_{\text{C}}[\phi_{\mathcal{D}_{\text{v}}\,e_1},\ldots,
  \phi_{\mathcal{D}_{\text{v}}\,e_{M_{\text{e}}}},
  \phi_{\mathcal{D}_{\text{v}}\,n}](x)}
  {\sqrt{\text{W}_{\text{C}}[\phi_{\mathcal{D}_{\text{v}}\,e_1},\ldots,
  \phi_{\mathcal{D}_{\text{v}}\,e_{M_{\text{e}}}}](x)
  \text{W}_{\text{C}}[\phi_{\mathcal{D}_{\text{v}}\,e_1},\ldots,
  \phi_{\mathcal{D}_{\text{v}}\,e_{M_{\text{e}}}}](x+1)}}\n
  &=\frac{\text{W}_{\text{C}}\Bigl[
  \frac{\text{W}_{\text{C}}[f_1,\ldots,f_l,u_1]}{w},\ldots,
  \frac{\text{W}_{\text{C}}[f_1,\ldots,f_l,u_m]}{w},
  \frac{\text{W}_{\text{C}}[f_1,\ldots,f_l,v]}{w}\Bigr](x)}
  {\sqrt{\text{W}_{\text{C}}\Bigl[
  \frac{\text{W}_{\text{C}}[f_1,\ldots,f_l,u_1]}{w},\ldots,
  \frac{\text{W}_{\text{C}}[f_1,\ldots,f_l,u_m]}{w}\Bigr](x)
  \text{W}_{\text{C}}\Bigl[
  \frac{\text{W}_{\text{C}}[f_1,\ldots,f_l,u_1]}{w},\ldots,
  \frac{\text{W}_{\text{C}}[f_1,\ldots,f_l,u_m]}{w}\Bigr](x+1)}}\n
  &\quad\times\frac{\prod_{j=0}^mG(x+j)}
  {\sqrt{\prod_{j=0}^{m-1}G(x+j)G(x+1+j)}},
  \label{rdQM:W/sqrtWW}
\end{align}
and a short calculation shows
\begin{align}
  &\quad\frac{\prod_{j=0}^mG(x+j)}{\sqrt{\prod_{j=0}^{m-1}G(x+j)G(x+1+j)}}\n
  &=\bigl((-1)^l\epsilon_{\mathcal{D}_{\text{v}}}\bigr)^{m+1}
  \biggl(\prod_{j=1}^lB(x+j-1)D(x+j)\cdot
  \prod_{j=m+1}^{l+m}B(x+j-1)D(x+j)
  \biggr)^{\frac18}.
  \label{rdQM:G/sqrtGG}
\end{align}
For the potential functions $B_{\mathcal{D}_{\text{v}}}(x)$ and
$D_{\mathcal{D}_{\text{v}}}(x)$ \eqref{rdQM:DDv},
a short calculation shows
\begin{align}
  &\quad\prod_{j=1}^m
  B_{\mathcal{D}_{\text{v}}}(x+j-1)
  D_{\mathcal{D}_{\text{v}}}(x+j)\n
  &=\biggl(\prod_{j=1}^mB(x+j-1)D(x+j)\cdot
  \prod_{j=l+1}^{l+m}B(x+j-1)D(x+j)
  \biggr)^{\frac12}\n
  &\quad\times\frac{
  \text{W}_{\text{C}}[f_1,\ldots,f_l](x)
  \text{W}_{\text{C}}[f_1,\ldots,f_l](x+m+1)}
  {\text{W}_{\text{C}}[f_1,\ldots,f_l](x+1)
  \text{W}_{\text{C}}[f_1,\ldots,f_l](x+m)}.
  \label{BDvDDvs}
\end{align}
{}From \eqref{BDvDDvs}, \eqref{rdQM:W/sqrtWW}--\eqref{rdQM:G/sqrtGG} and
\eqref{rdQM:Cas:id}, we obtain
\begin{align}
  \eqref{rdQM:phiDvDen}&=(-1)^{l+m}
  (-1)^{lm}\epsilon_{\mathcal{D}_{\text{v}}}\epsilon_{\mathcal{D}_{\text{e}}}
  \biggl(\prod_{j=1}^{l+m}B(x+j-1)D(x+j)\biggr)^{\frac14}\n
  &\quad\times
  \frac{\text{W}_{\text{C}}[f_1,\ldots,f_l,u_1,\ldots,u_m,v](x)}
  {\sqrt{\text{W}_{\text{C}}[f_1,\ldots,f_l,u_1,\ldots,u_m](x)
  \text{W}_{\text{C}}[f_1,\ldots,f_l,u_1,\ldots,u_m](x+1)}}\\
  &\stackrel{(\text{\romannumeral1})}{=}\eqref{rdQM:phiDn},
  \nonumber
\end{align}
namely the equality $\phi_{\mathcal{D}\,n}(x)=
\phi_{\mathcal{D}_{\text{v}}\mathcal{D}_{\text{e}}\,n}(x)$.
In (\romannumeral1) we have used $\epsilon_{\mathcal{D}}=(-1)^{lm}
\epsilon_{\mathcal{D}_{\text{v}}}\epsilon_{\mathcal{D}_{\text{e}}}$
because an ordered set $\mathcal{D}$ is now
$\{\text{v}_1,\ldots,\text{v}_{M_{\text{v}}},e_1,\dots,e_{M_{\text{e}}}\}$.

\section{Summary and Comments}
\label{sec:summary}

The Wronskian and Casoratian identities \eqref{Wformula}, \eqref{idQM:Cformula}
and \eqref{rdQM:Cformula} have played an important role in the study of
deformations of the quantum mechanical systems (oQM, idQM and rdQM,
respectively) by the multi-step Darboux transformations.
A generalization of the Wronskian identity \eqref{Wformula} is known as
Theorem\,\ref{thm:QM}.
Corresponding to this generalization, we have presented similar generalizations
of the Casoratian identities \eqref{idQM:Cformula} and \eqref{rdQM:Cformula}
as Theorem\,\ref{thm:idQM} and \ref{thm:rdQM}, respectively.

We have also discussed the application of these
Theorem\,\ref{thm:QM}--\ref{thm:rdQM} to quantum mechanical systems.
Multi-step Darboux transformations with both the virtual state wavefunctions
and the eigenstate wavefunctions as seed solutions are considered.
By interpreting this deformation in two ways, as \eqref{twodeform}, we obtain
two different expressions of the eigenfunctions.
The equality of these two expressions is shown by using
Theorem\,\ref{thm:QM}--\ref{thm:rdQM}.

The multi-indexed orthogonal polynomials $P_{\mathcal{D},n}$, whose
characteristic feature is the missing degrees, are obtained from the
eigenfunctions $\phi_{\mathcal{D}\,n}(x)$ by removing the ``ground state''
part \cite{gos,os22,os25,os26,os27,os35,idQMcH}.
The multi-indexed polynomials $P_{\mathcal{D}_{\text{v}},n}$ obtained from
\eqref{oQM:phiDvn}, \eqref{idQM:phiDvn} and \eqref{rdQM:phiDvn} are case-(1)
polynomials, namely the set of missing degrees $\mathbb{Z}_{\geq 0}\backslash
\{\deg P_{\mathcal{D}_{\text{v}},n}|n\in\mathbb{Z}_{\geq 0}\}$ is
$\{0,1,\ldots,\ell-1\}$.
For $\mathcal{D}_{\text{e}}\neq\emptyset$,
the multi-indexed polynomials $P_{\mathcal{D},n}$ obtained from
\eqref{oQM:phiDn}, \eqref{idQM:phiDn} and \eqref{rdQM:phiDn} are case-(2)
polynomials, namely the set of missing degrees is not $\{0,1,\ldots,\ell-1\}$.
Since the expressions \eqref{oQM:phiDn}, \eqref{idQM:phiDn} and
\eqref{rdQM:phiDn} are equal to \eqref{oQM:phiDvDen}, \eqref{idQM:phiDvDen}
and \eqref{rdQM:phiDvDen}, respectively, we obtain another expression of
$P_{\mathcal{D},n}$ from \eqref{oQM:phiDvDen}, \eqref{idQM:phiDvDen}
and \eqref{rdQM:phiDvDen}.
Namely, the case-(2) polynomials $P_{\mathcal{D},n}$ are expressed in terms
of the case-(1) polynomials $P_{\mathcal{D}_{\text{v}},n}$.
For their explicit forms, we leave them as an exercise for interested readers.

\section*{Acknowledgements}

I thank Ryu Sasaki for informing me of ref.\cite{swia} (He conjectured
Theorem\,\ref{thm:rdQM}), discussion and comments on the manuscript.
This work was supported by JSPS KAKENHI Grant Number JP19K03667.



\begin{thebibliography}{99}

\bibitem{os24}
S.\,Odake and R.\,Sasaki,
``Discrete quantum mechanics,'' (Topical Review)
J. Phys. {\bf A44} (2011) 353001 (47pp),
{\tt arXiv:1104.0473[math-ph]}.

\bibitem{gomez}
D.\,G\'{o}mez-Ullate, N.\,Kamran and R.\,Milson,
%
``An extended class of orthogonal polynomials defined by a
Sturm-Liouville problem,''
J. Math. Anal. Appl. {\bf 359} (2009) 352-367,
{\tt arXiv:0807.3939[math-\hspace{0pt}ph]}.

\bibitem{os16}
S.\,Odake and R.\,Sasaki,
``Infinitely many shape invariant potentials and new orthogonal polynomials,''
Phys. Lett. {\bf B679} (2009) 414-417,
{\tt arXiv:0906.0142[math-ph]}.

\bibitem{gos}
L.\,Garc\'ia-Guti\'errez, S.\,Odake and R.\,Sasaki,
``Modification of Crum's theorem for `discrete' quantum mechanics,''
Prog. Theor. Phys. {\bf 124} (2010) 1-26,
{\tt arXiv:1004.0289\hspace{0pt}[math-ph]}.

\bibitem{os22}
S.\,Odake and R.\,Sasaki,
``Dual Christoffel transformations,''
Prog. Theor. Phys. {\bf 126} (2011) 1-34,
{\tt arXiv:1101.5468[math-ph]}.

\bibitem{gomez2}
D.\,G\'{o}mez-Ullate, N.\,Kamran and R.\,Milson,
``Two-step Darboux transformations and exceptional Laguerre polynomials,"
J. Math. Anal. Appl. {\bf 387} (2012) 410-418,
{\tt arXiv:\hspace{0pt}1103.5724[math-ph]}.

\bibitem{os25}
S.\,Odake and R.\,Sasaki,
``Exactly solvable quantum mechanics and infinite families of
multi-indexed orthogonal polynomials,"
Phys. Lett. {\bf B702} (2011) 164-170,
{\tt arXiv:1105.\hspace{0pt}0508[math-ph]}.

\bibitem{os26}
S.\,Odake and R.\,Sasaki,
``Multi-indexed ($q$-)Racah polynomials,"
J. Phys. {\bf A 45} (2012) 385201 (21pp),
{\tt arXiv:1203.5868[math-ph]}.

\bibitem{os27}
S.\,Odake and R.\,Sasaki,
``Multi-indexed Wilson and Askey-Wilson polynomials,"
J. Phys. {\bf A46} (2013) 045204 (22pp),
{\tt arXiv:1207.5584[math-ph]}.

\bibitem{ggm13}
D.\,G\'{o}mez-Ullate, Y.\,Grandati and R.\,Milson,
``Rational extensions of the quantum harmonic oscillator and exceptional
Hermite polynomials,''
J. Phys. {\bf A47} (2014) 015203 (27pp),
{\tt arXiv:1306.5143[math-ph]}.

\bibitem{d14}
A.\,J.\,Dur\'{a}n,
``Exceptional Meixner and Laguerre orthogonal polynomials,"
J. Approx. Theory {\bf 184} (2014) 176-208,
{\tt arXiv:1310.4658[math.CA]}.

\bibitem{d17}
A.\,J.\,Dur\'{a}n,
``Exceptional Hahn and Jacobi orthogonal polynomials,"
J. Approx. Theory {\bf 214} (2017) 9-48,
{\tt arXiv:1510.02579[math.CA]}.

\bibitem{os35}
S.\,Odake and R.\,Sasaki,
``Multi-indexed Meixner and Little $q$-Jacobi (Laguerre) Polynomials,"
J. Phys. {\bf A50} (2017) 165204 (23pp),
{\tt arXiv:1610.09854[math.CA]}.

\bibitem{idQMcH}
S.\,Odake,
``Exactly Solvable Discrete Quantum Mechanical Systems and Multi-indexed
Orthogonal Polynomials of the Continuous Hahn and Meixner-Pollaczek Types,''
Prog. Theor. Exp. Phy. {\bf 2019} (2019) 123A01 (20pp),
{\tt arXiv:1907.12218[math-ph]}.

\bibitem{swia}
M.\,Swiatkowski,
``Wronskian Identities,''
Pi Mu Epsilon J. {\bf 5} (1971) 191-194.

\bibitem{os29}
S.\,Odake and R.\,Sasaki,
``Krein-Adler transformations for shape-invariant potentials and pseudo
virtual states,"
J. Phys. {\bf A46} (2013) 245201 (24pp),
{\tt arXiv:1212.6595[math-\hspace{0pt}ph]}.

\bibitem{gkm11} 
D.\,G\'{o}mez-Ullate, N.\,Kamran and R.\,Milson,
``On orthogonal polynomials spanning a non-standard flag,''
Contemp. Math. {\bf 563} (2011) 51-72,
{\tt arXiv:1101.5584[math-ph]}.

%
\bibitem{krein}
M.\,G.\,Krein,
``On continuous analogue of a formula of Christoffel from the theory
of orthogonal polynomials," (Russian)
Doklady Acad. Nauk. CCCP, {\bf 113} (1957) 970-973.

\bibitem{adler}
V.\,\'E.\,Adler,
``A modification of Crum's method,''
Theor. Math. Phys. {\bf 101} (1994) 1381-1386.

\bibitem{rrmiop3}
S.\,Odake,
``Recurrence Relations of the Multi-Indexed Orthogonal Polynomials : $\III$,''
J. Math. Phys. {\bf 57} (2016) 023514 (24pp),
{\tt arXiv:1509.08213[math-ph]}.

\end{thebibliography}
\end{document}